\documentclass[aps,pre,twocolumn,superscriptaddress]{revtex4-1}
\usepackage{amsmath}
\usepackage{amssymb}
\usepackage{graphicx}
\usepackage{color}

\begin{document}

\title{Autocorrelation study of the $\Theta$ transition for a coarse-grained polymer model}

\author{Kai Qi}
\email{kaiqi@physast.uga.edu}
\affiliation{Soft Matter Systems Research Group, Center for Simulational
Physics, The University of Georgia, Athens, GA 30602, USA}
\author{Michael Bachmann}
\email{bachmann@smsyslab.org; http://www.smsyslab.org}
\affiliation{Soft Matter Systems Research Group, Center for Simulational
Physics, The University of Georgia, Athens, GA 30602, USA}
\affiliation{Instituto de F\'{\i}sica,
Universidade Federal de Mato Grosso, 78060-900 Cuiab\'a (MT), Brazil}
\affiliation{Departamento de F\'{\i}sica,
Universidade Federal de Minas Gerais, 31270-901 Belo Horizonte (MG), Brazil}
\date{\today}

\begin{abstract}
By means of Metropolis Monte Carlo simulations of a coarse-grained
model for flexible polymers, we investigate how the integrated autocorrelation
times of different energetic and structural quantities depend on the
temperature. We show that, due to critical slowing down, an extremal
autocorrelation time can also be considered as an indicator for the collapse
transition that helps to locate the transition point. This is particularly
useful for finite systems, where response quantities such as the specific heat
do not necessarily exhibit clear indications for pronounced thermal activity.
\end{abstract}

\maketitle


\section{INTRODUCTION}

The necessity for a better understanding of general physical principles
and mechanisms of structural transitions of polymers, such as folding,
crystallization, aggregation, and the adsorption at solid and soft substrates
has provoked numerous computational studies of polymer models. Autocorrelation
properties of such models govern the statistical accuracy of estimated
expectation values of physical quantities but also help illustrate the dynamic
behavior or the relaxation properties.
Verdier and co-workers were among the
first to investigate autocorrelations of a simple lattice polymer
approach, in which the Brownian motion of the monomers
is simulated by kinetic displacements of single monomers
\cite{Verdier1962,Verdier1966,Verdier1972,Verdier1973}. By using Monte Carlo
methods, the autocorrelation functions and relaxation times of structural
quantities were calculated in order to study dynamic properties of random-coil
polymer chains such as the relaxation of asphericity in lattice-
model chains with and without excluded volume interaction
\cite{Kranbuehl1973,Kranbuehl1977}. More recently, these studies were extended
to continuous models, where autocorrelation properties of the center-of-mass
velocity, Rouse coordinates, end-to-end distance, end-to-end vector, normal
modes, and the radius of gyration for polymer melts
\cite{McCormick2005,Pestryaev2011,Aoyagi2001}, and of dynamic quantities of a
polymer immersed in a solution
\cite{Malevanets2000,Polson2006,Bishop1979,Rapaport1979,Bruns1981,Bansal1981,
Mussawisade2005} were investigated. Integrated autocorrelation times are also
employed to judge the efficiency of importance-sampling algorithms
\cite{Nidras1997}. However, much less is known about how autocorrelation times
and structural transitions of polymers depend on each other.

In the past, most of the studies on analyzing the properties of the
autocorrelation times focused on spin models. The second-order phase
transition between ferromagnetism and paramagnetism is characterized by a
divergent spatial correlation length $\xi$ at the transition point $T_c$. In
the thermodynamic limit (i.e., infinite system size), the divergent behavior
is given by $\xi \sim \epsilon^{-\nu}$, where $\epsilon \equiv |1-T/T_c|$ and
$\nu$ is a critical exponent \cite{Landau2000,NewmanMC1999,WJanke2002}. If an
importance sampling Monte Carlo method is employed
\cite{Landau2000,WJanke1996,WJanke1998}, the number of configurational updates
that is needed to decorrelate the information about the history of macroscopic
system states is measured by the autocorrelation time $\tau$. It is described
by the power law $\tau \propto \xi^z \propto \epsilon^{-\nu z}$, where $z$
denotes the dynamic critical exponent, which depends on the employed algorithm
\cite{Landau2000,NewmanMC1999,WJanke2002}. However, in a 
system of finite size, the correlation length can never really diverge. This
is because the largest possible cluster has the volume $L^d$, where $L$ is the
system size and $d$ is the dimensionality. Thus, the divergence of the
correlation length as well as the autocorrelation time are ``cut off'' at the
boundary, i.e., $\xi \lesssim L$. Consequently, $\tau \sim L^z$ at
temperatures sufficiently close to the critical point
\cite{Landau2000,NewmanMC1999,WJanke2002}. For local updates, such as single
spin flips, and by using the Metropolis algorithm \cite{MetroRosTell1953}, the
autocorrelation time becomes very large near the critical temperature because
the dynamic critical exponent is in this case $z\approx2$. This effect is
usually called critical slowing down, but it can be reduced significantly if
non-local updates, such as in Swendsen-Wang, Wolff, and multigrid algorithms
\cite{NewmanMC1999,WJanke1998,Sokal1989,Sokal1992}, are employed. Metropolis
simulations with local updates yield for the Ising model 
$z\approx 2.1665$ in 2D and $z \approx 2.02 $ in 3D
\cite{NewmanMC1999,Nightingale1996,Matz1994}. For non-local updates, numerical
estimates yield a $z$ value less than unity
\cite{Coddington1992,NewmanMC1999,Kandel1988,Kandel1989}.

Since most phase transitions in nature are of first order
\cite{Gunton1983,Binder1987,Herrmann1992,JankeFirst-Order1994},
it is also useful to discuss autocorrelation properties
near first-order phase transitions. In
a finite system, the characteristic feature of a first-order transition is the
double-peaked energy distribution with an entropic suppression regime
between the two peaks.
The dip is caused by the entropic contribution to the Boltzmann factor $\propto \text{exp}(-2
\sigma L^{d-1})$, where $\sigma$ is the (reduced) interface tension and
$L^{d-1}$ is the projected area of the interfaces. Thus, the dynamics in a
canonical ensemble will suffer from
the ``supercritical slowing down'', in which the tremendous average
residence time the system spends in a pure phase
is described by the autocorrelation time $\tau \propto \text{exp}(2\sigma L^{d-1})$
\cite{WJanke2002,JankePhaseTransitions}. Since this slowing down is related to the
shape of the energetic probability distribution itself, it is impossible to reduce the
autocorrelation time by using cluster or multigrid algorithms. The simulation
in a generalized ensemble, such as multicanonical ensemble, where the slowing
down can be reduced to a
powerlike behavior with $\tau \propto L^{d\alpha}$ ($\alpha \approx 1$)
\cite{Berg1991,Janke1992,Berg1993,Billoire1993,Grossmann1992,Janke1993},
can overcome this difficulty.

In this paper, we will investigate autocorrelation properties of
different quantities for elastic, flexible polymers, described by a simple
coarse-grained model. The thermodynamic behavior of the system is simulated by
local monomer displacement and Metropolis Monte Carlo sampling, resembling
Brownian dynamics in a canonical ensemble. The goal is to identify structural
transitions and transition temperatures for this model.

The paper is structured as follows. The coarse-grained polymer model,
the simulation method,
and a brief introduction to autocorrelation theory
are described in Sec.~\ref{model_methods}. 
Simulation results are presented and discussed in Sec.~\ref{results}. 
Our conclusions are summarized in Sec.~\ref{summary}.


\section{MODEL AND METHODS}
\label{model_methods}


\subsection{Model}

For our study, we use a generic model of a single flexible, elastic
polymer chain \cite{MBachmann}. Monomers adjacent in the linear chain are
bonded by the anharmonic FENE (finitely extensible nonlinear elastic)
potential \cite{BiCuArHa1987, MilBhaBin2001}
\begin{equation}
	V_{\text{FENE}}(r_{ii+1})=-\frac{K}{2}R^2 \text{ln}\left[1-\left(\frac{r_{ii+1}-r_0}{R}\right)^2\right]. \label{FENE}
\end{equation}
We set $r_0=1$, which represents the distance where the FENE potential
is minimum, $R=3/7$, and $K=98/5$. Non-bonded monomers interact via a
truncated, shifted Lennard-Jones potential
\begin{equation}
	V_{\text{LJ}}^{\text{mod}}(r_{ij})= V_{\text{LJ}}(r_{ij}) -  V_{\text{LJ}}(r_{c}), \label{LJ1}
\end{equation}
with
\begin{equation}
	V_{\text{LJ}}(r_{ij})= 4\epsilon \left[ \left( \frac{\sigma}{r_{ij}} \right)^{12} - \left( \frac{\sigma}{r_{ij}} \right)^{6} \right], \label{LJ2}
\end{equation}
where we choose the energy scale to be $\epsilon=1$ and the length scale to be
$\sigma=r_0/2^{1/6}$. The cut-off radius is set to $r_c=2.5\sigma$ so that
$V_{\text{LJ}}(r_{c}) \approx -0.0163\epsilon$. For $r_{ij}>r_c$,
$V_{\text{LJ}}^{\text{mod}}(r_{ij}) \equiv 0$. The total energy of a
conformation $\zeta = (\vec{r}_1,\cdots,\vec{r}_{L})$ for a chain with $L$
monomers reads 
\begin{equation}
	E(\zeta) = \sum^{L-2}_{i=1} \sum^{L}_{j=i+2} V_{\text{LJ}}^{\text{mod}}(r_{ij}) + \sum^{L-1}_{i=1} V_{\text{FENE}}(r_{ii+1}). \label{totalE}
\end{equation}


\subsection{Simulation Method}

In our simulations, we employed the Metropolis Monte Carlo method. In a single
MC update, the conformation is changed by a random local displacement of a
monomer. Once a monomer is randomly chosen, its position is changed within a
small cubic box with edge lengths $d=0.3r_0$. Denoting the inverse thermal
energy by $\beta = 1/k_{\mathrm{B}} T$, with $k_{\mathrm{B}} \equiv 1$ in our
simulations, the probability of accepting such an update is given by the
Metropolis criterion~\cite{MetroRosTell1953}
\begin{equation}
	p = \text{min}(1,\text{exp}[-\beta(E_{\text{new}}-E_{\text{old}})]), \label{Metro.Crit.}
\end{equation}
where $E_{\text{old}}$ and $E_{\text{new}}$ are the energies before and after
the proposed update. According to Eq.~\eqref{Metro.Crit.}, an update will be
directly accepted if $E_{\text{new}}\le E_{\text{old}}$. If $E_{\text{new}} >
E_{\text{old}}$, the update will be accepted only with the probability
$e^{-\beta \delta E}$, where $\delta E = E_{\text{new}}-E_{\text{old}}$. In
each simulation, we performed about $3 \times 10^8$ sweeps after extensive
equilibration. A sweep contains $L$ Monte Carlo steps, where $L$ is the number
of monomers for a chosen polymer.


\subsection{Autocorrelation Theory}


Suppose a time series with a large number of data from an importance sampling
MC simulation has been generated, the expectation value of any quantity $O$
can be estimated by calculating the arithmetic mean over the Markov chain,
\begin{equation}
	\overline{O}=\frac{1}{N}\sum_{j=1}^{N} O_j , \label{mean}
\end{equation}
where $O_j$ is the value of $O$ in the $j$th measurement and $N$ is the
number of total  measurements. In equilibrium, the expectation value of
$\overline{O}$ is the same as the expectation value of the individual
measurement 
\begin{equation}
	\langle \overline{O} \rangle = \frac{1}{N}\sum_{j=1}^{N} \langle O_j \rangle = \langle O \rangle, \label{mean_expectation}
\end{equation}
because of time-translational invariance. In Metropolis simulations, the
individual measurements will not be independent. Thus, by introducing the
normalized autocorrelation function $(A(0)=1)$,
\begin{equation}
	A(k)=\frac{\langle O_l O_{l+k}\rangle - \langle O_l \rangle^2}{\sigma_O^2},  \label{auto_func}
\end{equation}
where $l$ can be any integer in the range $\left[ 1, N-k \right]$ and
$\sigma_O^2=\langle O_l^2 \rangle - \langle O_l \rangle^2 = \langle O^2
\rangle - \langle O \rangle^2$, the corresponding variance of $\overline{O}$
is calculated as \cite{WJanke2002} 
\begin{eqnarray}
	\sigma^2_{\overline{O}} &=& \langle \overline{O}^2 \rangle -\langle \overline{O} \rangle^2 \nonumber \\
				&=& \frac{2\sigma_O^2}{N} \left[\frac{1}{2} + \sum_{k=1}^{N} A(k) \left( 1-\frac{k}{N} \right) \right]. \label{mean_var_1}
\end{eqnarray}
For large time separation $k$, the autocorrelation function decays exponentially,
\begin{equation}
	A(k) \longrightarrow e^{-k/\tau_{O,\text{exp}}}, \label{tau_exp}
\end{equation}
where $\tau_{O,\text{exp}}$ is the exponential autocorrelation time of $O$.
Because of large statistical fluctuations in the tail of $A(k)$, the accurate
estimation of $\tau_{O,\text{exp}}$ is often difficult. By introducing the
integrated autocorrelation time, 
\begin{equation}
	\tau'_{O,\text{int}} = \frac{1}{2} + \sum_{k=1}^{N} A(k) \left( 1-\frac{k}{N} \right), \label{tau_int}
\end{equation}
Eq.~\eqref{mean_var_1} becomes 
\begin{equation}
	\epsilon^2_{\overline{O}} \equiv \sigma^2_{\overline{O}} = \frac{2\sigma_O^2}{N} \tau'_{O,\text{int}} = \frac{\sigma_O^2}{N_{\text{eff}}} \label{error}
\end{equation}
with the effective statistics $N_{\text{eff}} =
N/2\tau'_{O,\text{int}}$. According to Eq.~\eqref{tau_exp}, in any meaningful
simulation with $N\gg\tau_{O,\text{exp}}$, we can safely neglect the
correction term in the parentheses in Eq.~\eqref{tau_int}. This leads to the
frequently employed definition of the integrated autocorrelation time,
\begin{equation}
	\tau_{O,\text{int}} = \frac{1}{2} + \sum_{k=1}^N A(k). \label{tau_int_u}
\end{equation} 
The estimation of the integrated autocorrelation time requires the
replacement of the expectation value in $A(k)$ by mean values, e.g., $\langle
O_l O_{l+k} \rangle$ and $\langle O_l \rangle $ by $\overline{O_l O_{l+k}} $
and $\overline{O}_l$. Therefore, it is useful to introduce the following
estimator
\begin{equation}
	\tilde{\tau}_{O,\text{int}}(k_{\text{max}}) = \frac{1}{2} +\sum_{k=1}^{k_{\text{max}}} \tilde{A}(k) \label{tau_int_est}
\end{equation}
where $\tilde{A}(k)$ is the estimator of $A(k)$. Since $\tilde{A}(k)$ usually
decays to zero as $k$ increases, $\tilde{\tau}_{O,\mathrm{int}}$ will finally
converge to a constant. Because of the statistical noise of $\tilde{A}(k)$ for
large $k$, $\tilde{\tau}_{O,\text{int}}$ is obtained by averaging
$\tilde{A}(k)$ over several independent runs.


The standard estimator for the variance of $O$ is
\begin{equation}
	\tilde{\sigma}_O^2 =\overline{O^2} - \overline{O}^2 = \overline{(O-\overline{O})^2} = \frac{1}{N}\sum_{i=1}^{N} (O_i - \overline{O})^2, \label{var_est}
\end{equation}
and its expected value is
\begin{equation}
	\langle \tilde{\sigma}_O^2 \rangle = \langle  \overline{O^2} - \overline{O}^2 \rangle
					   = \sigma^2_O \left(1-\frac{1}{N_{\text{eff}}} \right),  \label{var_est_exp}
\end{equation}
with $\sigma_O^2=\langle O^2 \rangle - \langle O \rangle^2$. It is obvious
that this form systematically underestimates the true value by a term of the
order of $\tau_{O,\text{int}}/N$. The $2\tau_{O,\text{int}}/N$ correction is
the systematic error due to the finiteness of the time series, and it is
called bias. Even in the case in which all the data are uncorrelated
($\tau_{O,\text{int}}=1/2$), the estimator is still biased, $\langle
\tilde{\sigma}_O^2 \rangle=\sigma^2_O \left(1 - 1/N  \right)$. Thus, it is
reasonable to define the bias-corrected estimator
\begin{equation}
	\tilde{\sigma}^2_{O,\text{c}} \equiv
\frac{N_{\text{eff}}}{N_{\text{eff}}-1}\tilde{\sigma}^2_{O} =
\frac{1}{N-2\tau_{O,\text{int}}}\sum_{i=1}^{N} (O_i - \overline{O})^2,
\label{var_est_bias}
\end{equation}
which satisfies $\langle \tilde{\sigma}^2_{O,\text{c}}
\rangle=\sigma^2_O$. Thus, the bias-corrected estimator for the squared error
of the mean value becomes 
\begin{equation}
	\epsilon^2_{\overline{O}} = \frac{\tilde{\sigma}^2_{O,\text{c}}}{N_{\text{eff}}} = \frac{1}{N(N_{\text{eff}}-1)}\sum_{i=1}^N \left(
 O_i - \overline{O} \right)^2. \label{error_corr}
\end{equation} 
For uncorrelated data, the error formula simplifies to 
\begin{equation}
	\epsilon^2_{\overline{O}} = \frac{\tilde{\sigma}^2_{O,\text{c}}}{N} = \frac{1}{N(N-1)}\sum_{i=1}^N \left(
 O_i - \overline{O} \right)^2. \label{error_corr_uncorrelated}
\end{equation}
%
%
%
%
\begin{figure*}
	\centering
	\includegraphics[width =7.8cm]{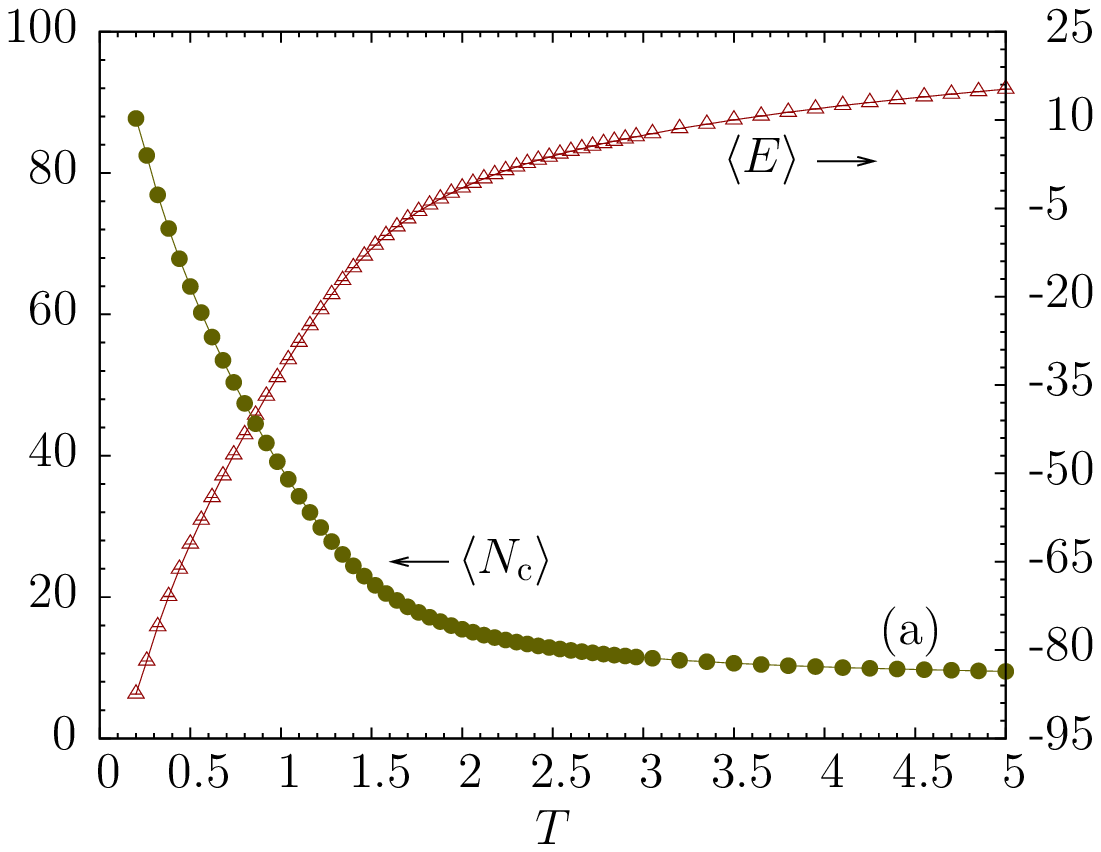}
	\includegraphics[width =7.8cm]{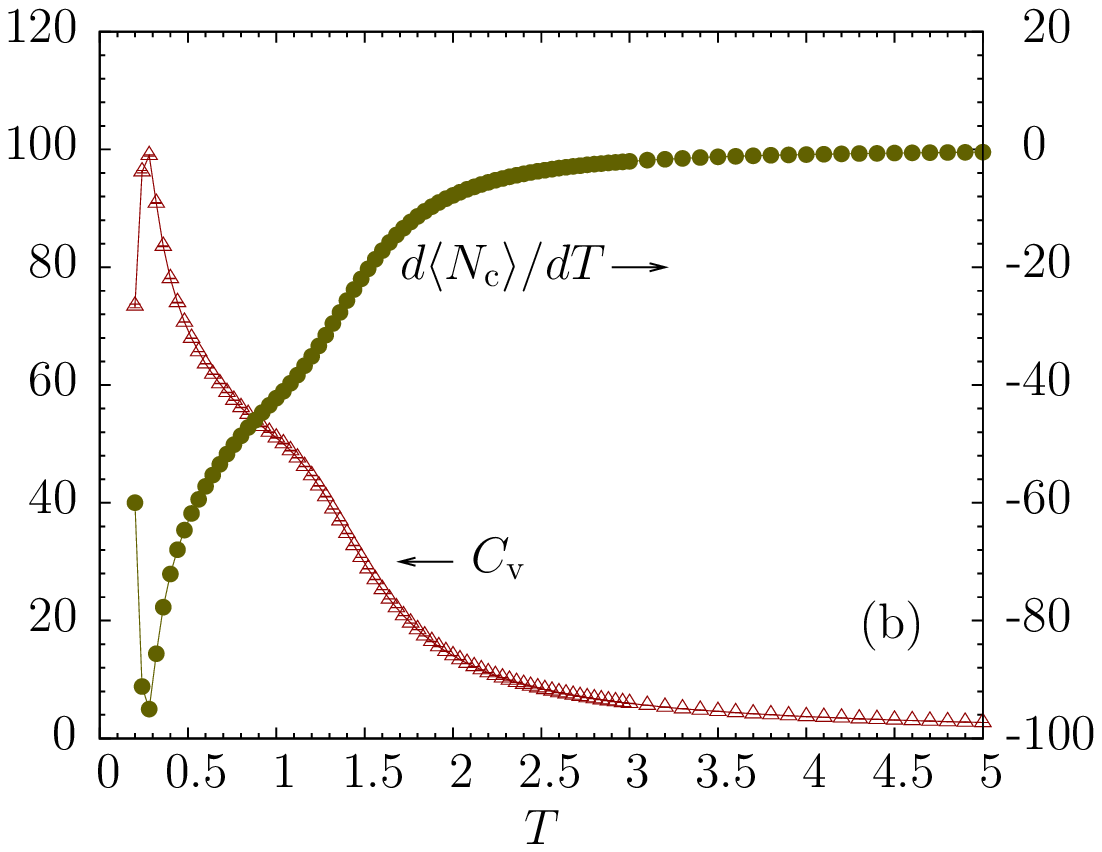}
	\includegraphics[width =7.8cm]{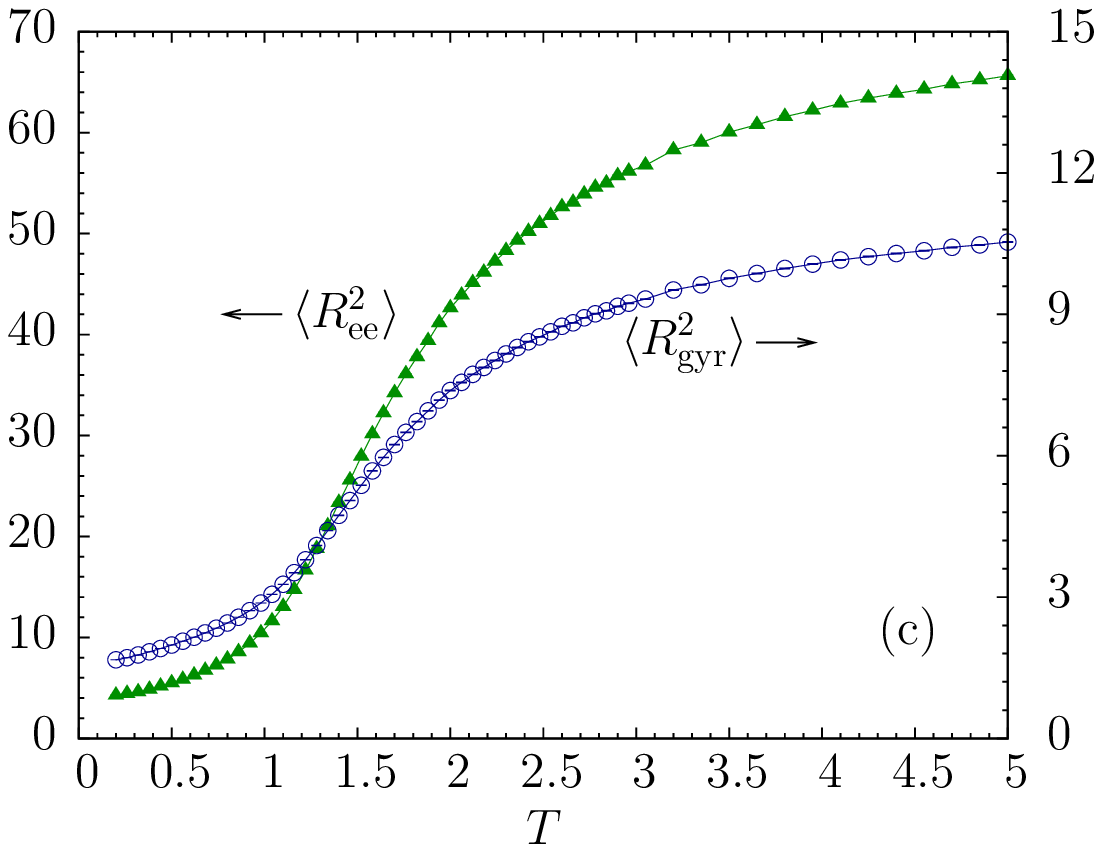}
	\includegraphics[width =7.8cm]{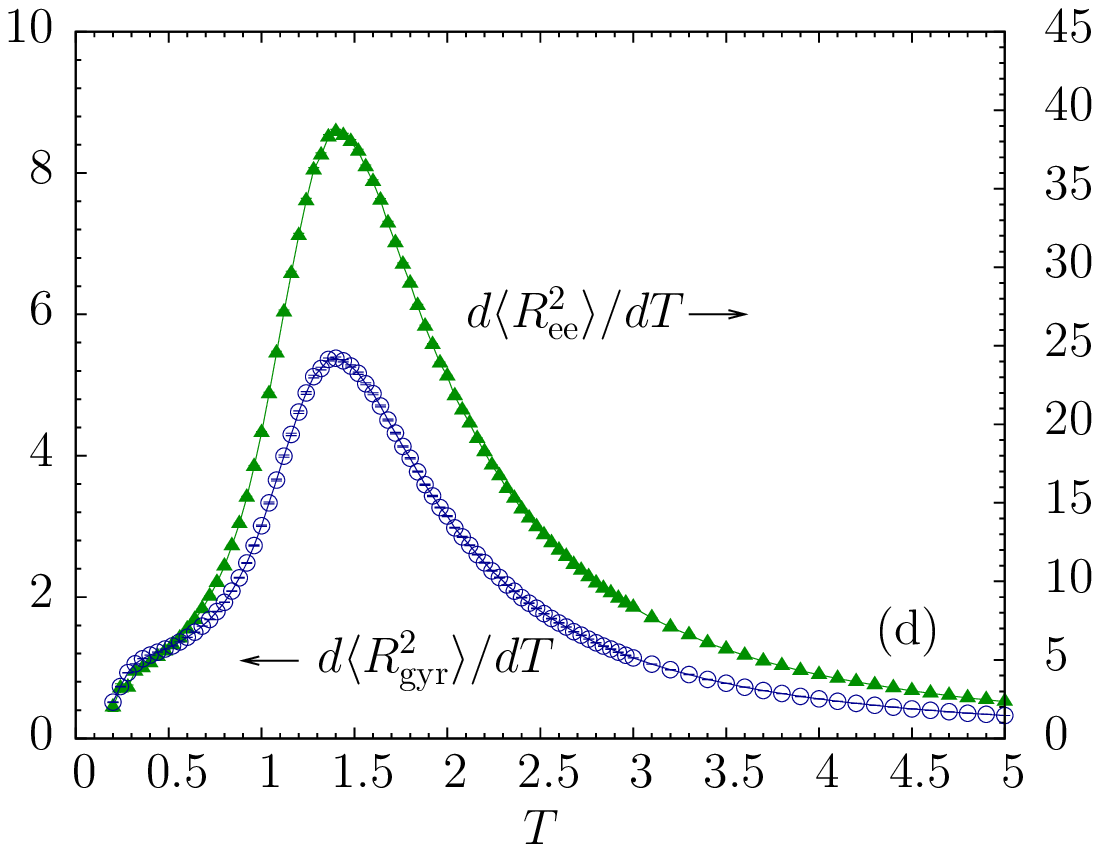}
	\caption{\label{canonical_quantities} (a) Mean energy $\langle E
\rangle$ and number of contacts $\langle N_{\mathrm{c}} \rangle$; (b) heat
capacity $C_{\text{V}}$ and thermal fluctuation of the number of contacts
$d\langle N_{\mathrm{c}} \rangle / dT$; (c) square end-to-end distance
$\langle R^2_{\mathrm{ee}} \rangle$ and square radius of gyration $\langle
R^2_{\mathrm{gyr}} \rangle$; (d) thermal fluctuations of the square end-to-end
distance $d \langle R^2_{\mathrm{ee}} \rangle / dT$ and the square radius of
gyration $d \langle R^2_{\mathrm{gyr}} \rangle / dT$ for a flexible polymer
with 30 monomers. Error bars are smaller than the symbol size.}
\end{figure*}

Integrated autocorrelation times can also be estimated by using the so-called
binning method. Assuming that the time series consists of $N$ correlated
measurements $O_i$, this time series can be divided into $K$ bins, which
should be large enough so that the correlation of the data in each bins decays
sufficiently ($N_{\mathrm{B}} \gg \tau_{O,\text{int}}$). In this way, a set of
$K$ uncorrelated data subsets is generated, each of which contains
$N_{\mathrm{B}}$ data points such that $N=N_{\mathrm{B}} K$. The binning block
average $\overline{O}^{\mathrm{B}}_k$ of the $k$-th block is calculated as 
\begin{equation}
	\overline{O}^{\mathrm{B}}_k= \frac{1}{N_{\mathrm{B}}} \sum_{i=1}^{N_{\mathrm{B}}} O_{(k-1)N_{\mathrm{B}}+i},\quad k=1,\ldots,K, \label{bin_avg} 
\end{equation}
and 
\begin{equation}
	\overline{O} = \frac{1}{K} \sum_{k=1}^K \overline{O}^{\mathrm{B}}_k\, , \label{bin_avg_total}
\end{equation}
coincides with the average \eqref{mean}. Since each bin average represents an
independent measurement, the variance of the binning block averages
$\sigma^2_{\overline{O}^{\mathrm{B}}}$ can be estimated from
Eq.~\eqref{var_est_bias}, 
\begin{equation}
	\tilde{\sigma}^2_{\overline{O}^{\mathrm{B}},\mathrm{c}} = \frac{1}{K-1} \sum_{k=1}^K \left( \overline{O}^{\mathrm{B}}_k - \overline{O} \right)^2, \label{var_bin}
\end{equation}  
and the statistical error of the mean value $\epsilon^2_{\overline{O}} \equiv \sigma^2_{\overline{O}} = \sigma^2_{\overline{O}^{\mathrm{B}}}/K$ is given by 
\begin{equation}
	\epsilon^2_{\overline{O}} = \frac{\tilde{\sigma}^2_{\overline{O}^{\mathrm{B}},\mathrm{c}}}{K} = \frac{1}{K(K-1)} \sum_{k=1}^K \left(\overline{O}^{\mathrm{B}}_k -\overline{O}\right)^2. \label{error_bin}
\end{equation}
By comparing this expression with Eq.~\eqref{error} and considering
Eqs.~\eqref{tau_int} and \eqref{tau_int_u}, we see that
$\sigma^2_{\overline{O}^{\mathrm{B}}}/K=2\tau_{O,\text{int}}\sigma^2_{O}/N$.
Hence, the autocorrelation time can also be estimated by means of the
binning variance as 
\begin{equation}
	\tilde{\tau}_{O,\text{bin}} =\frac{1}{2} N_{\mathrm{B}} \frac{\tilde{\sigma}^2_{\overline{O}^{\mathrm{B}},\mathrm{c}}}{\tilde{\sigma}^2_O}. \label{tau_bin}
\end{equation}
Since the bin averages are supposed to be uncorrelated, we utilized the
standard estimator \eqref{var_est} for the variance of the individual
measurements $\sigma^2_O$ ($N \gg 2\tau_{O,\text{int}}$). This method is more
convenient than the integration method \eqref{tau_int_est} since a precise
estimate of the autocorrelation function is not needed. For uncorrelated data
and $N_{\mathrm{B}}=1$,
$\tilde{\sigma}^2_{\overline{O}^{\mathrm{B}},\mathrm{c}}=\tilde{\sigma}^2_O$
for $N \gg 1$. Consequently
$\tilde{\tau}_{O,\text{bin}}=\tau_{O,\text{int}}=1/2$. In the correlated case,
too small bin sizes will underestimate the autocorrelation time. Given a time
series consisting of $N$ measurements, the estimator $\tilde{\sigma}^2_O$
remains unchanged if $N_{\mathrm{B}}$ is modified. Increasing $N_{\mathrm{B}}$
reduces the number of bins $K$ which leads to the decrease of the variance
$\tilde{\sigma}^2_{\overline{O}^{\mathrm{B}},\mathrm{c}}$. However, the
decrease rate is not the same as $N_{\mathrm{B}}$ is increased. Thus, the
right hand 
side of \eqref{tau_bin} will converge to a constant value identical to
$\tau_{O,\text{int}}$. Therefore, one typically plots the right hand side of
Eq.~\eqref{tau_bin} for various values of $N_{\mathrm{B}}$ and estimates
$\tau_{O,\text{int}}$ by reading the value the curve converges to
\cite{WJanke2002,MBachmann}.
\begin{figure*}
	\centering
	\includegraphics[width =5.9cm]{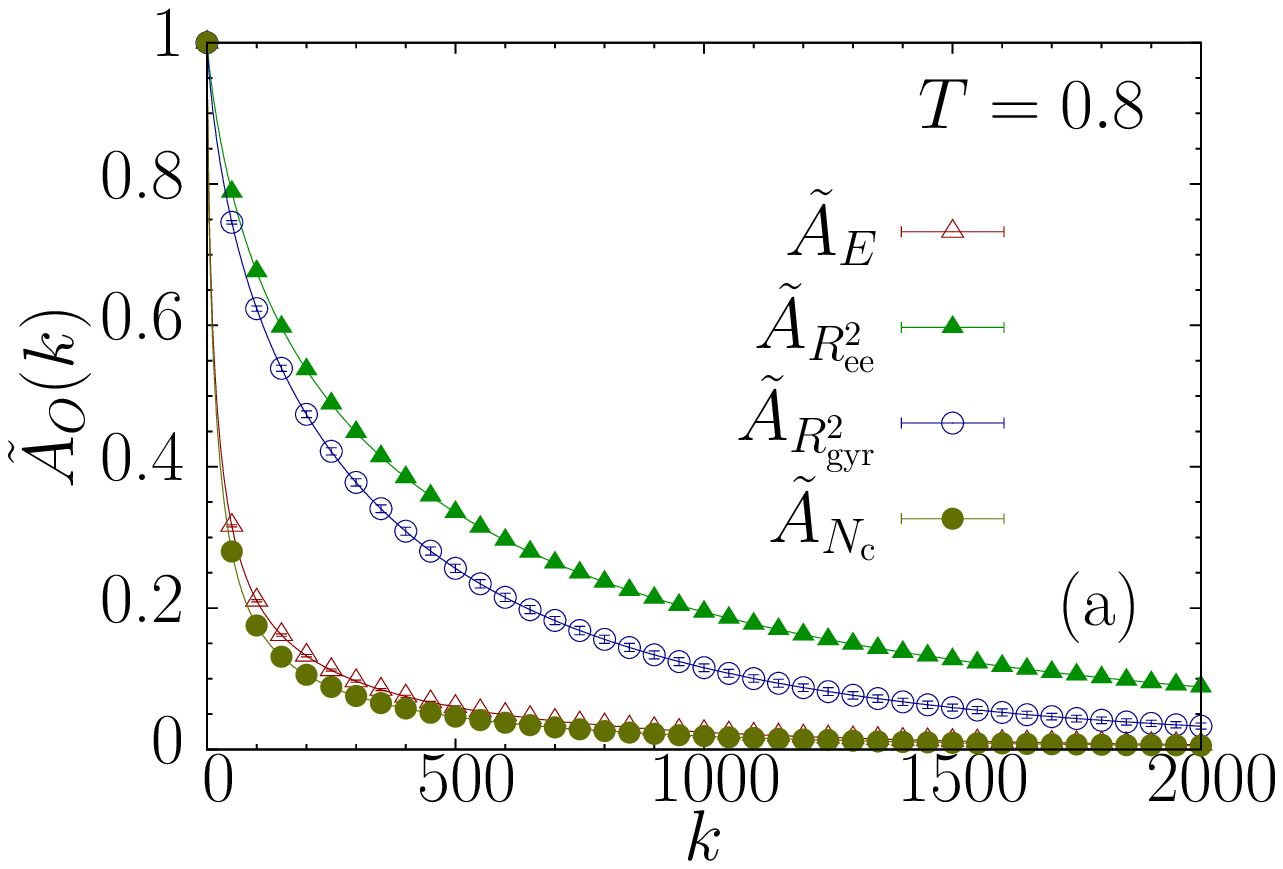}
	\includegraphics[width =5.6cm]{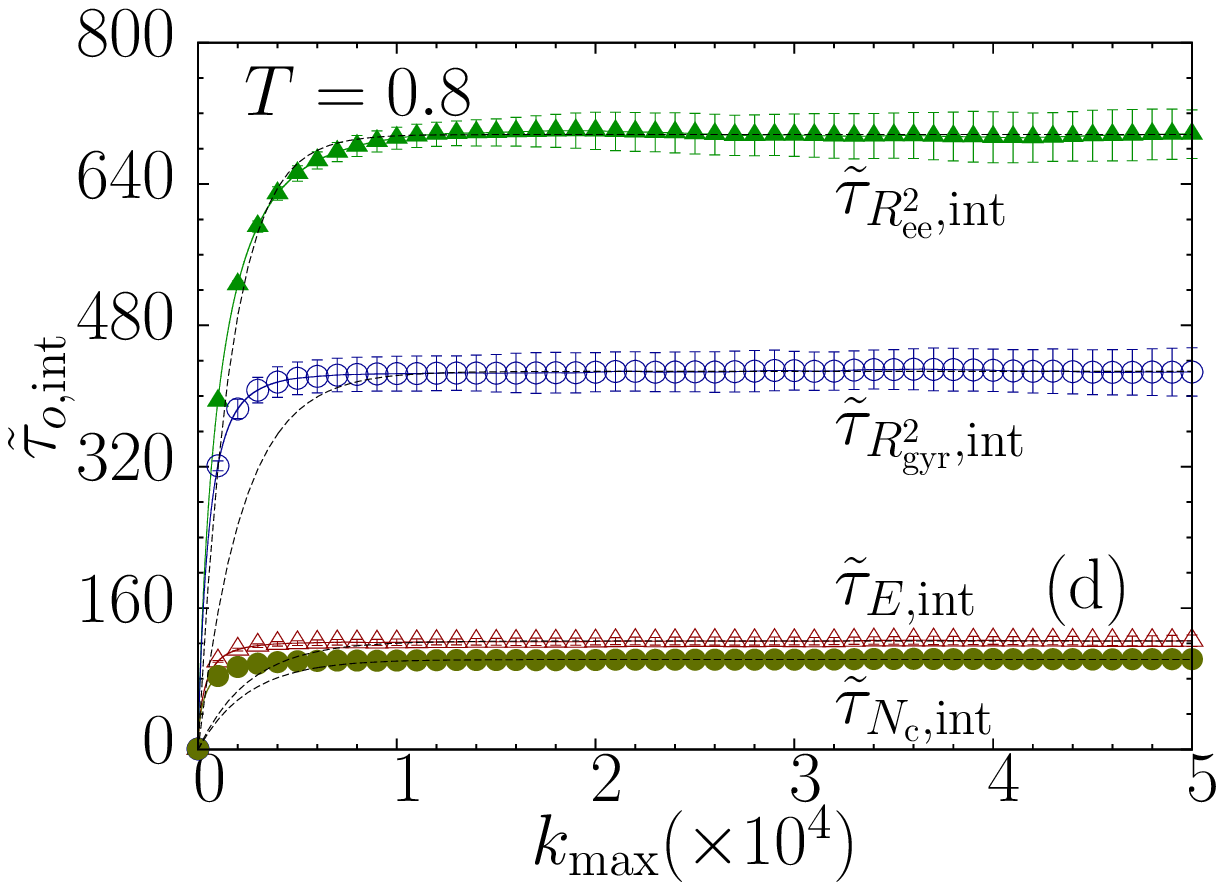}
	\includegraphics[width =5.6cm]{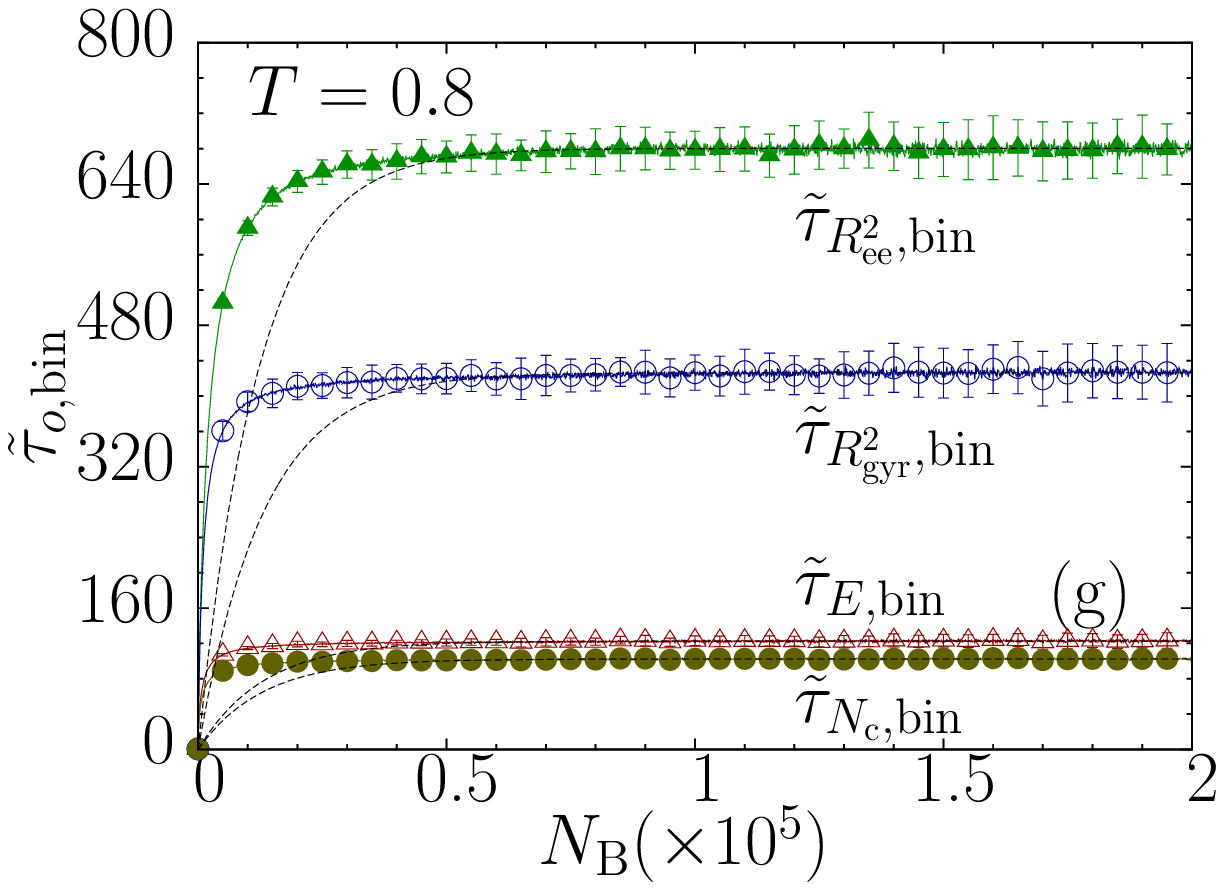}
	\includegraphics[width =5.9cm]{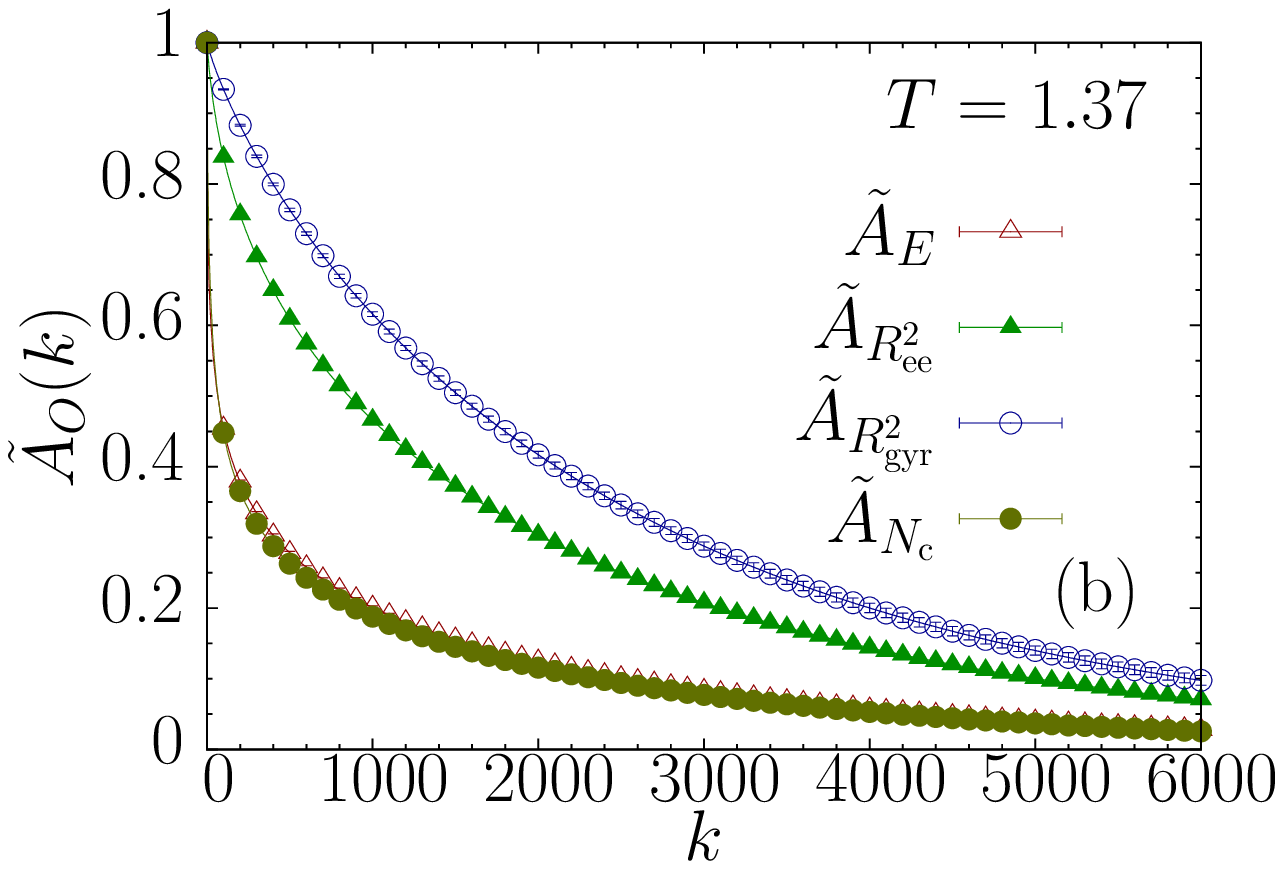}
	\includegraphics[width = 5.6cm]{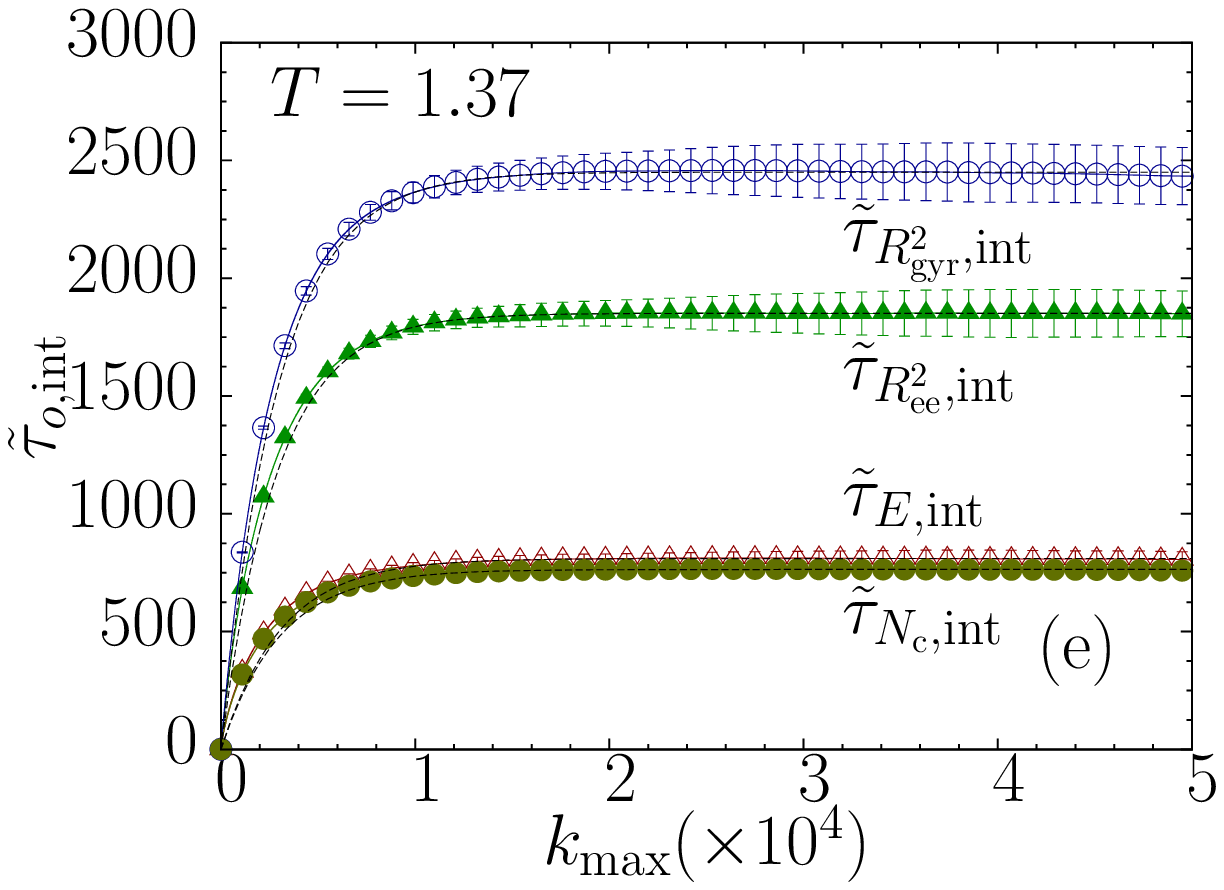}
	\includegraphics[width = 5.6cm]{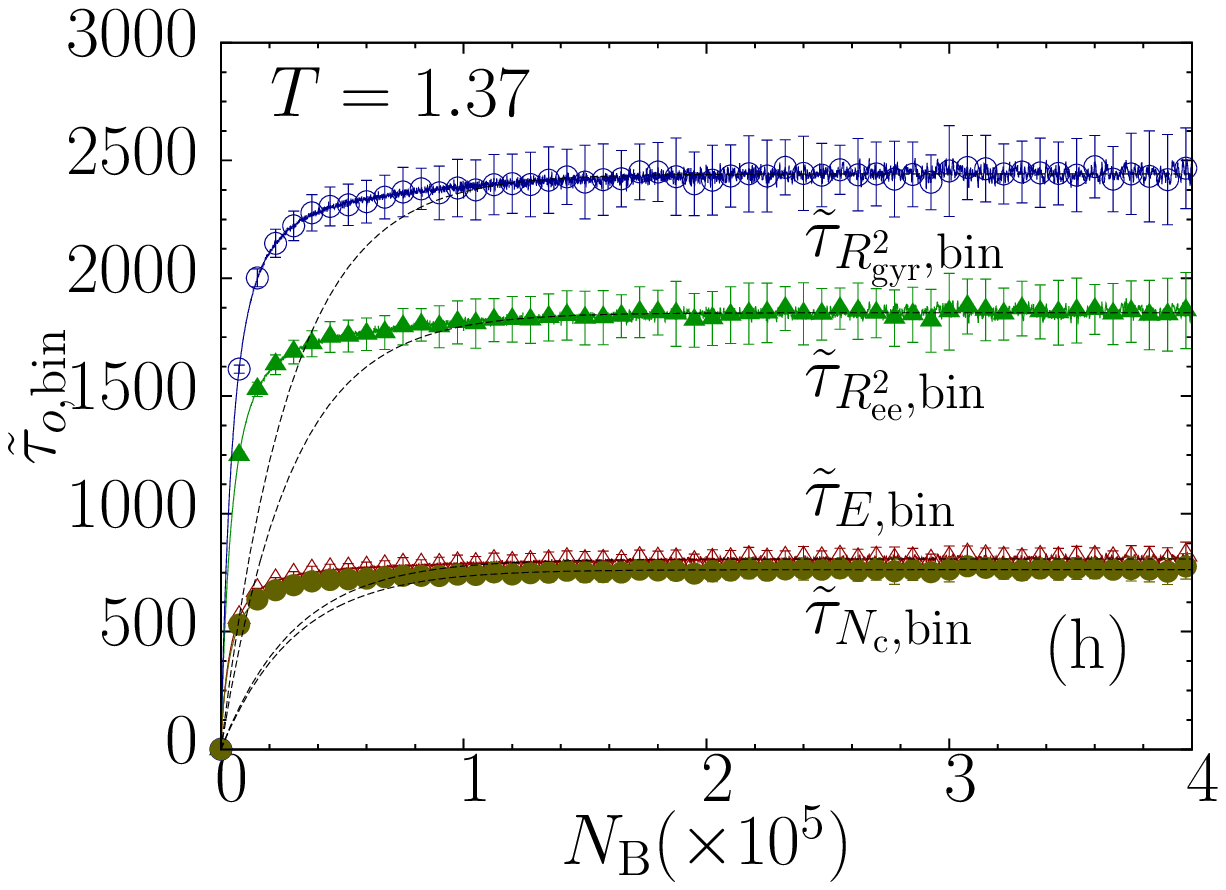}
	\includegraphics[width = 5.9cm]{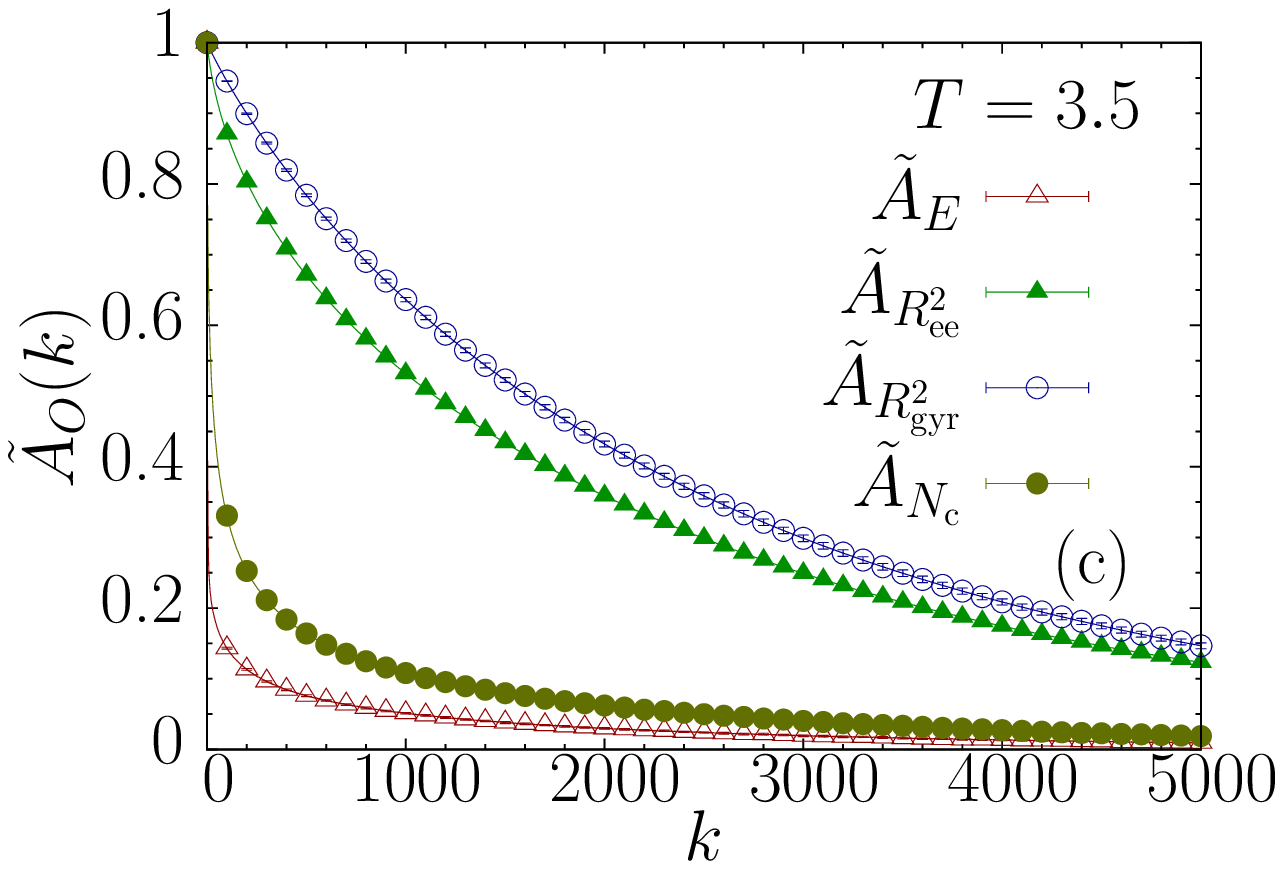}
        \includegraphics[width = 5.6cm]{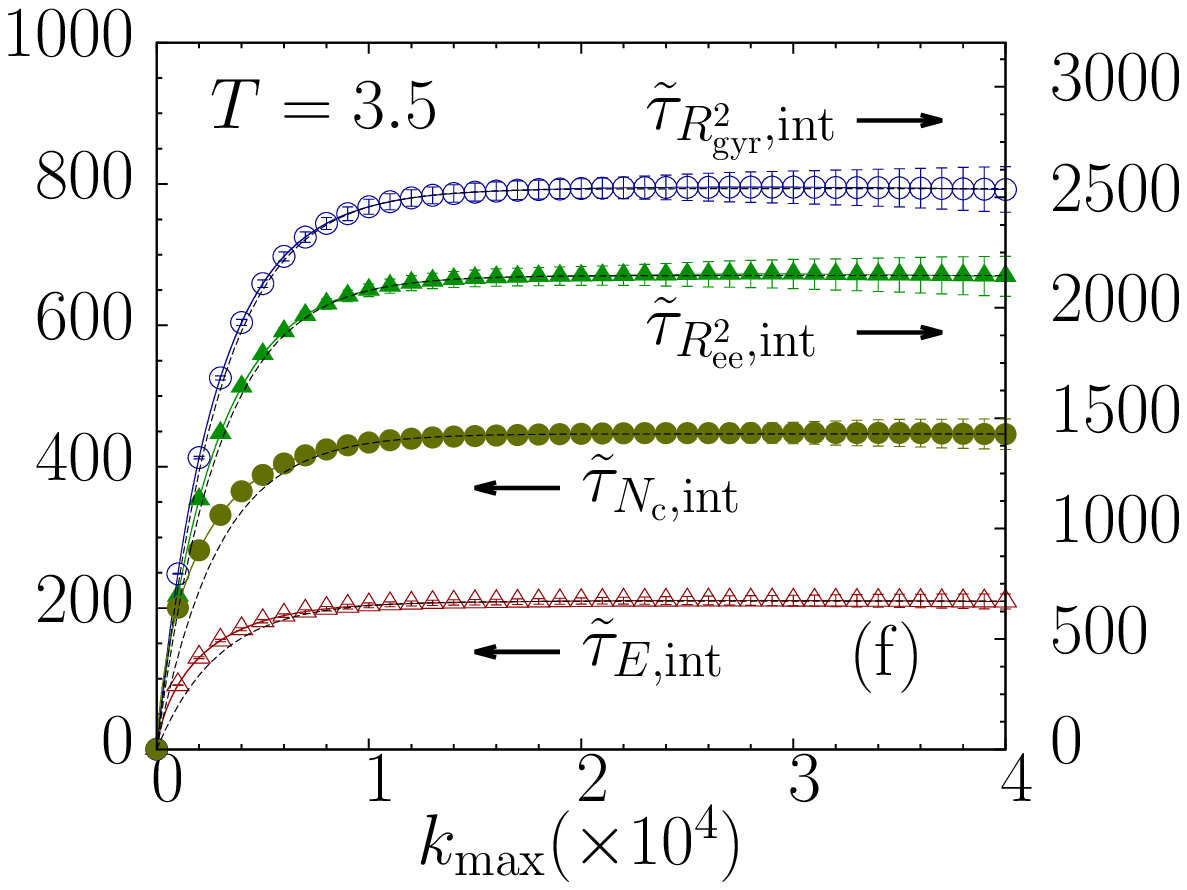}
	\includegraphics[width = 5.6cm]{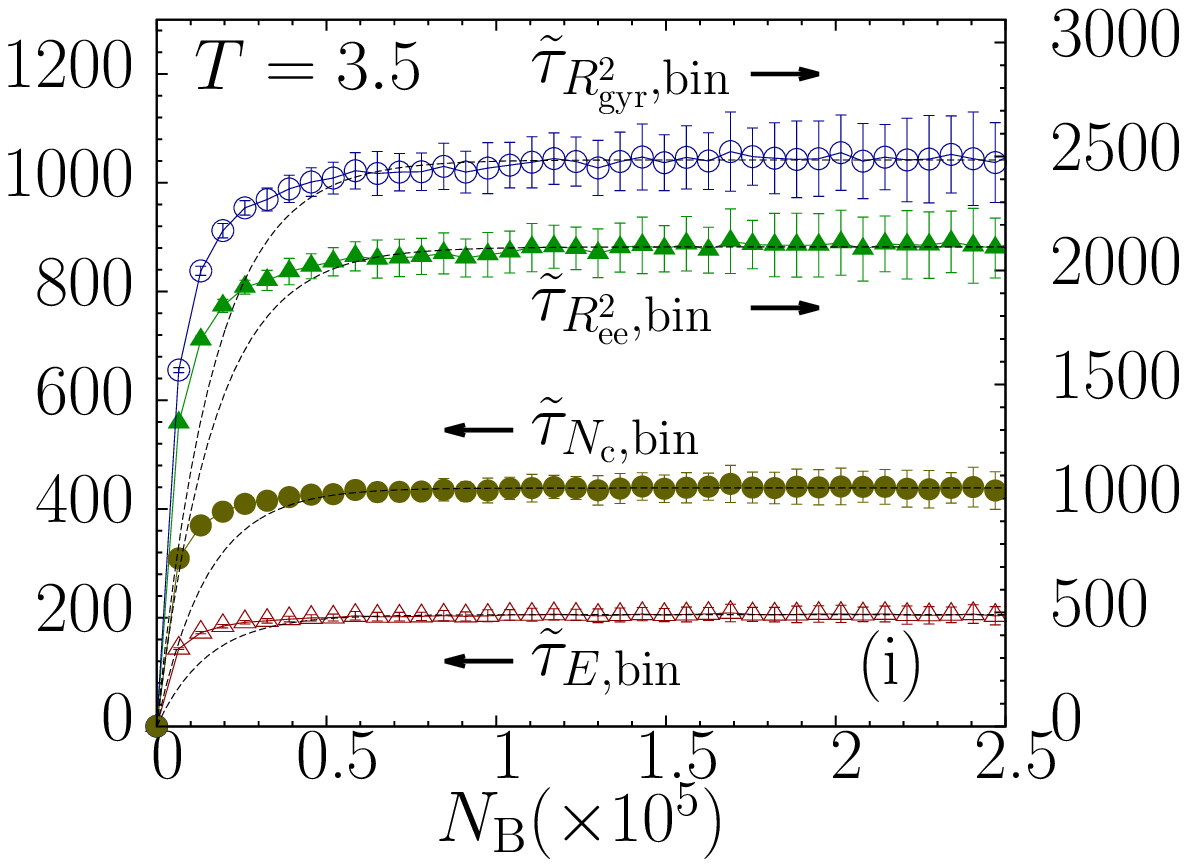}
	\caption{\label{auto_30_seperate} (a), (b), and (c)
Autocorrelation
functions of $E$, $R^2_{\mathrm{ee}}$, $R^2_{\mathrm{gyr}}$, and
$N_{\mathrm{c}}$ at temperatures below, near, and above the collapse
transition temperature, respectively, for $L=30$. For each quantity, the
estimated integrated autocorrelation time converges to a constant as shown in
(d), (e), and (f). The corresponding binning analysis results also show
good convergence and are plotted in (g), (h), and (i). Dashed lines
represent the fitted curves. Values of the fitted autocorrelation times the
curves converge to are listed in Table~\ref{table1}.}
\end{figure*}
%
%
%

\section{RESULTS}
\label{results}


\subsection{Autocorrelation times at constant displacement}

For the interpretation of the autocorrelation times of energy,
square end-to-end distance, square radius of gyration, and number of contacts,
it is helpful to first investigate thermodynamic properties of these
quantities. We have plotted the mean values of energy and number of contacts
in Fig.~\ref{canonical_quantities}(a), as well as the heat capacity and
thermal fluctuation of the number of contacts in
Fig.~\ref{canonical_quantities}(b). A contact between two non-bonded monomers
is formed, if their distance is in the interval $r_{ij}\in [0.8,1.2]$ for the
30-mer and $r_{ij}\in[0.87,1.13]$ for the 55-mer.
The number of contacts is a simple discrete order parameter which is also
helpful in distinguishing phases. It has proven to be particularly useful in
studies of lattice models \cite{MBachmann2006-1,MBachmann2006-2,TVogel2007}.
In the continuous model used here, it is a robust parameter that does not depend
on energetic model details.
Square end-to-end distance
and square radius of gyration curves are shown in
Fig.~\ref{canonical_quantities}(c) and their thermal fluctuations in
Fig.~\ref{canonical_quantities}(d). The two clear peaks at $T \approx 1.4$ of
the latter represent the collapse transition of the 30-mer. Note that the
fluctuations of energy and contact number in
Fig.~\ref{canonical_quantities}(b) do not exhibit a peak at the 
transition point, but only a ``shoulder''. As the temperature decreases,
dissolved or random coils (gas phase) collapse in a cooperative arrangement of
the monomers, and compact globular conformations (liquid phase) are favorably
formed. As the temperature decreases further, the polymer transfers from the
globular phase to the ``solid'' phase which is characterized by locally
crystalline or amorphous metastable structures. A corresponding peak and
valley which mark the liquid-solid (crystallization) or freezing transition of
the 30-mer can be observed at $T \approx 0.28$ in the heat capacity and
$d\langle N_{\mathrm{c}} \rangle / dT$ curves, respectively, in
Fig.~\ref{canonical_quantities}(b). These results coincide qualitatively with
those of a previous study, where a slightly different model was employed
\cite{Schnabel2011}. Due to insufficient Metropolis sampling at low
temperatures, we did not include data in the $T<0.2$ region.

\begin{table*}
\caption{Autocorrelation times of $E$, $R^2_{\mathrm{ee}}$,
$R^2_{\mathrm{gyr}}$, and $N_{\mathrm{c}}$ estimated by integration of
autocorrelation functions and by using the binning method at three
temperatures below, near, and above the collapse transition.\label{table1}} 

\begin{tabular}{|@{\hspace{3pt}}c@{\hspace{3pt}}||@{\hspace{6pt}}c@{\hspace{
6pt}}|@{\hspace{6pt}}c@{\hspace{6pt}}||@{\hspace{6pt}}c@{\hspace{6pt}}|@{
\hspace{6pt}}c@{\hspace{6pt}}||@{\hspace{6pt}}c@{\hspace{6pt}}|@{\hspace{6pt}}
c@{\hspace{6pt}}||@{\hspace{6pt}}c@{\hspace{6pt}}|@{\hspace{6pt}}c@{\hspace{
6pt}}|}
\hline
$T$ & $\tilde{\tau}_{E,\text{int}}$ & $\tilde{\tau}_{E,\text{bin}}$ &
$\tilde{\tau}_{N_{\text{c}},\text{int}}$ &
$\tilde{\tau}_{N_{\text{c}},\text{bin}}$ &
$\tilde{\tau}_{R^2_{\text{ee}},\text{int}}$ &
$\tilde{\tau}_{R^2_{\text{ee}},\text{bin}}$ &
$\tilde{\tau}_{R^2_{\text{gyr}},\text{int}}$ &
$\tilde{\tau}_{R^2_{\text{gyr}},\text{bin}}$ \\ 
\hline
\enspace 0.8 \enspace & 122 $\pm$ 7\enspace & 122 $\pm$ 13 & $101 \pm 7
\enspace$ & 102 $\pm$ 13 & 696 $\pm$ 33 & 680 $\pm$ 75 & 427 $\pm$ 28 & 426
$\pm$ 50 \\
\hline
1.37 & 810 $\pm$ 45 & 808 $\pm$ 94 & 763 $\pm$ 39 & 763 $\pm$ 93 & 1851 $\pm$
103 & 1853 $\pm$ 201 & 2450 $\pm$ 138 & 2443 $\pm$ 272 \\
\hline
3.5\enspace  & 209 $\pm$ 13 & 205 $\pm$ 25 & 446 $\pm$ 27 & 438 $\pm$ 52 &
2145 $\pm$ 106 & 2103 $\pm$ 228 & 2539 $\pm$ 121 & 2485 $\pm$ 268 \\ 
\hline
\end{tabular}
\end{table*}

\begin{figure*}
	\centering
	\includegraphics[width = 7.8cm]{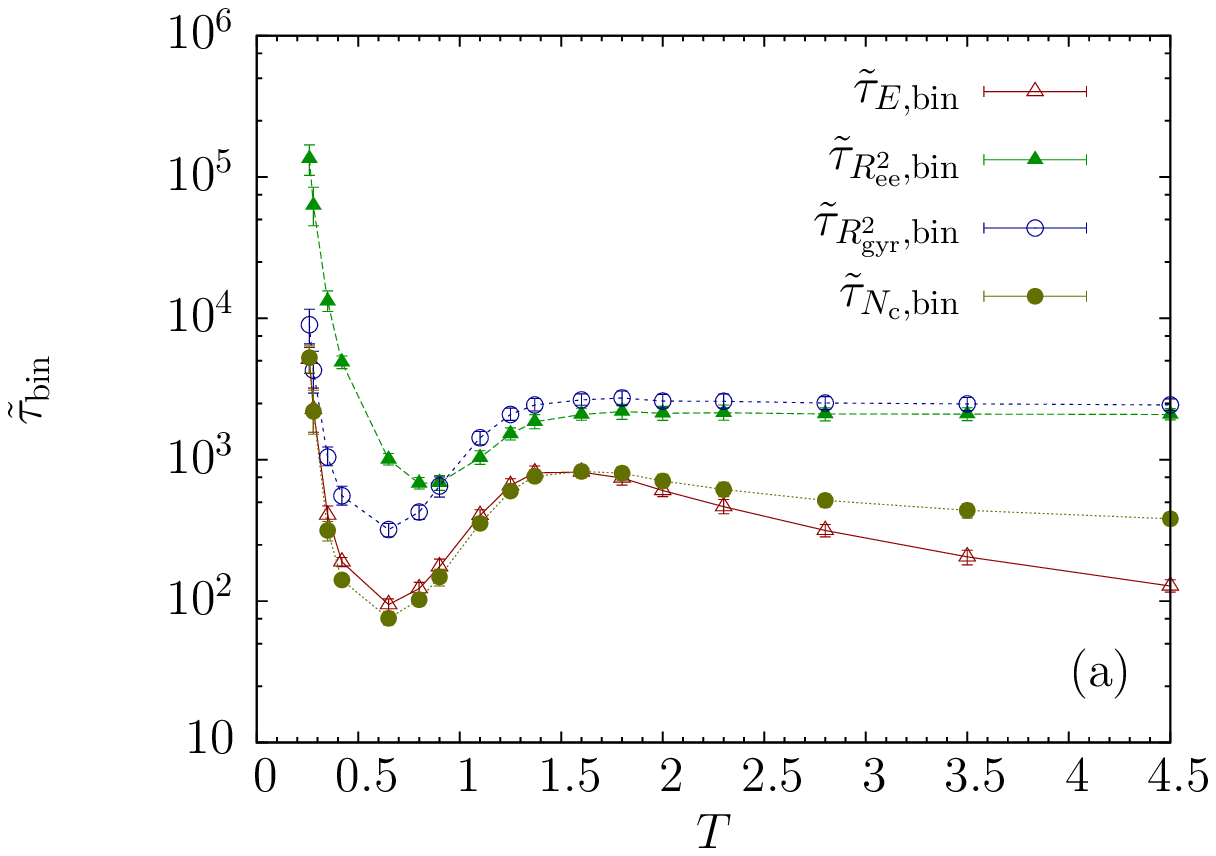}
	\includegraphics[width = 7.8cm]{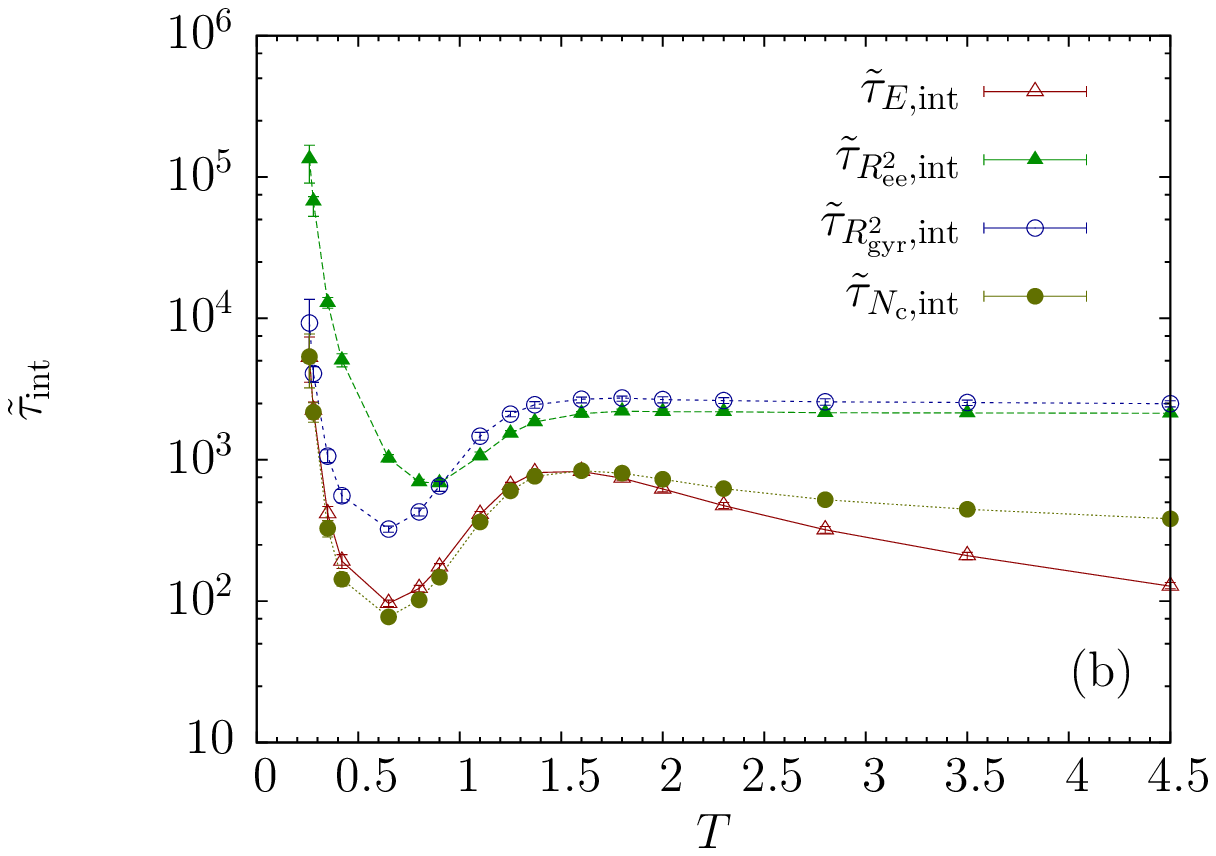} 
	\caption{\label{auto_30} Temperature dependence of
integrated autocorrelation times (a) estimated with the binning method; (b)
obtained by the integration of autocorrelation functions for the 30-mer.}
\end{figure*}

We performed the integration of the autocorrelation \eqref{tau_int_est} and
the binning analysis to estimate the integrated autocorrelation times at $17$
temperatures in the interval $T \in [0.26,4.5]$ for the 30-mer and at $16$
temperatures in the interval $T\in[0.3,5]$ for the 55-mer. Mean values
$\overline{Q}_O(x)$ (where $\overline{Q}_O(x)$ stands for
$\overline{\tilde{A}}_O(k)$,
$\overline{\tilde{\tau}}_{O,\text{bin}}(N_{\mathrm{B}})$ or
$\overline{\tilde{\tau}}_{O,\text{int}}(k_{\text{max}})$) for a quantity $O$
were calculated at each temperature in $N_{\mathrm{r}}$ ($N_{\mathrm{r}}>20$)
independent runs:
\begin{equation}
	\overline{Q}_O(x)=\frac{1}{N_{\mathrm{r}}} \sum_{i=1}^{N_{\mathrm{r}}} Q_O^i (x), \label{V_mean}
\end{equation}
where $Q_O^i (x)$ is the value calculated in the $i$th run. As shown in
Fig.~\ref{auto_30_seperate}, all estimates of autocorrelation functions and
times converge for large values of $k$, $k_{\text{max}}$, and
$N_{\mathrm{B}}$, respectively, as expected. The error of $\overline{Q}_O(x)$
is estimated by 
\begin{equation}
	\epsilon^2_{\overline{Q}_O(x)} = \frac{1}{N_{\mathrm{r}}-1} \sum_{i=1}^{N_{\mathrm{r}}} \left( Q^i_O(x) - \overline{Q}_O(x) \right)^2, \label{V_error} 
\end{equation} 
because all runs were performed independently of each other. The consistency
of the two different methods used for the estimation of autocorrelation times
for the investigated quantities become apparent from Table~\ref{table1}, where
we have listed the autocorrelation time estimate for three temperatures below,
near, and above the $\Theta$ point. The results coincide within the numerical
error bars.

In order to estimate the integrated autocorrelation time systematically, we
performed least-squares fitting for all the curves in both the integration
method of the autocorrelation function and binning analysis at each
temperature. The empirical fit function for any quantity $O$ is chosen to be
of the form
\begin{equation}
	f_O (x) = \tau^f_O (1-e^{-x/x^f}), \label{fit_func} 
\end{equation}
where $x$ represents $k_{\text{max}}$ in the integration of the
autocorrelation functions method and $N_{\mathrm{B}}$ in binning analysis;
$\tau^f_O$ and $x^f$ are two fit parameters. The fitting curves, also plotted
in Fig.~\ref{auto_30_seperate}, coincide well with the mean values of the
integrated autocorrelation times in the $N_{\mathrm{B}}/k_{\text{max}}$
region, where convergence sets in.

\begin{figure*}
	\centering
	\includegraphics[width=5.8cm]{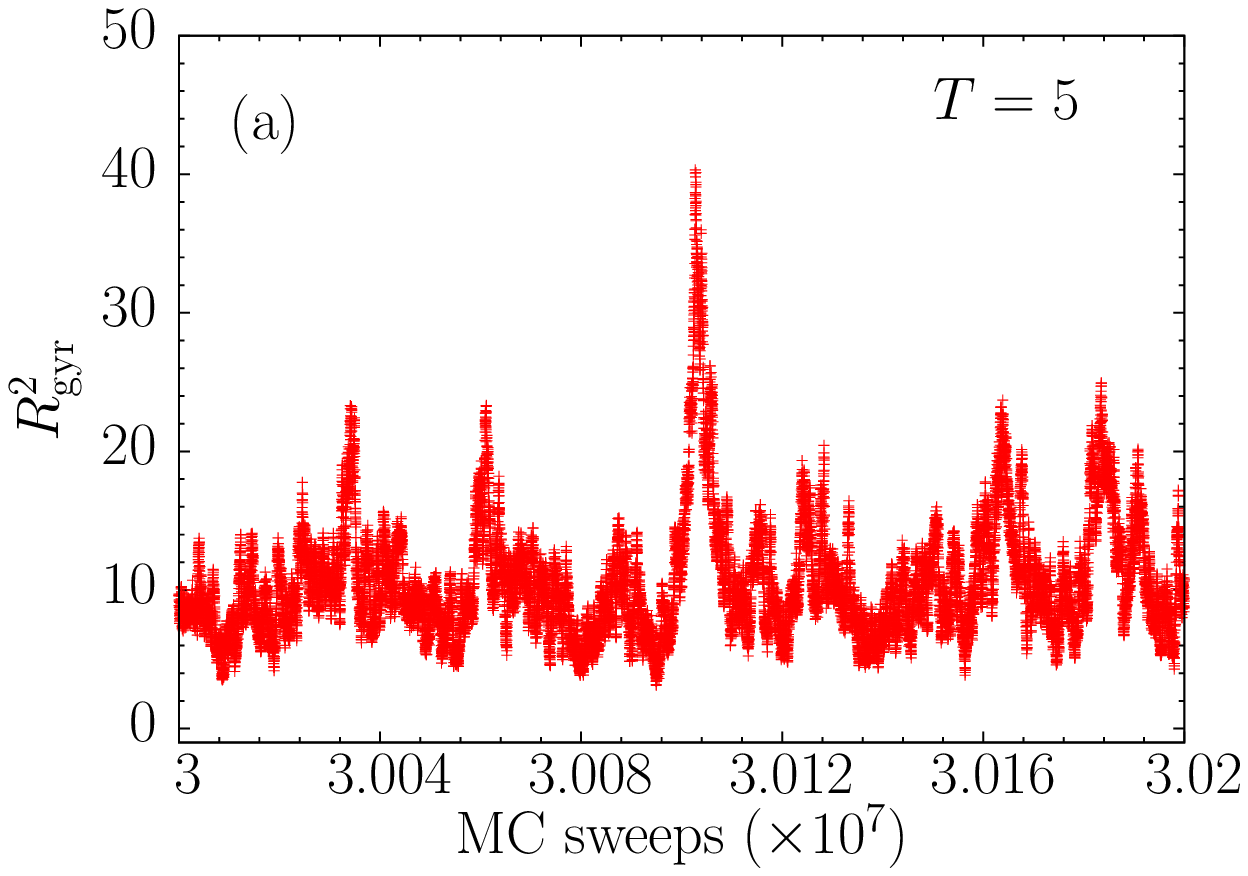}
	\includegraphics[width=5.8cm]{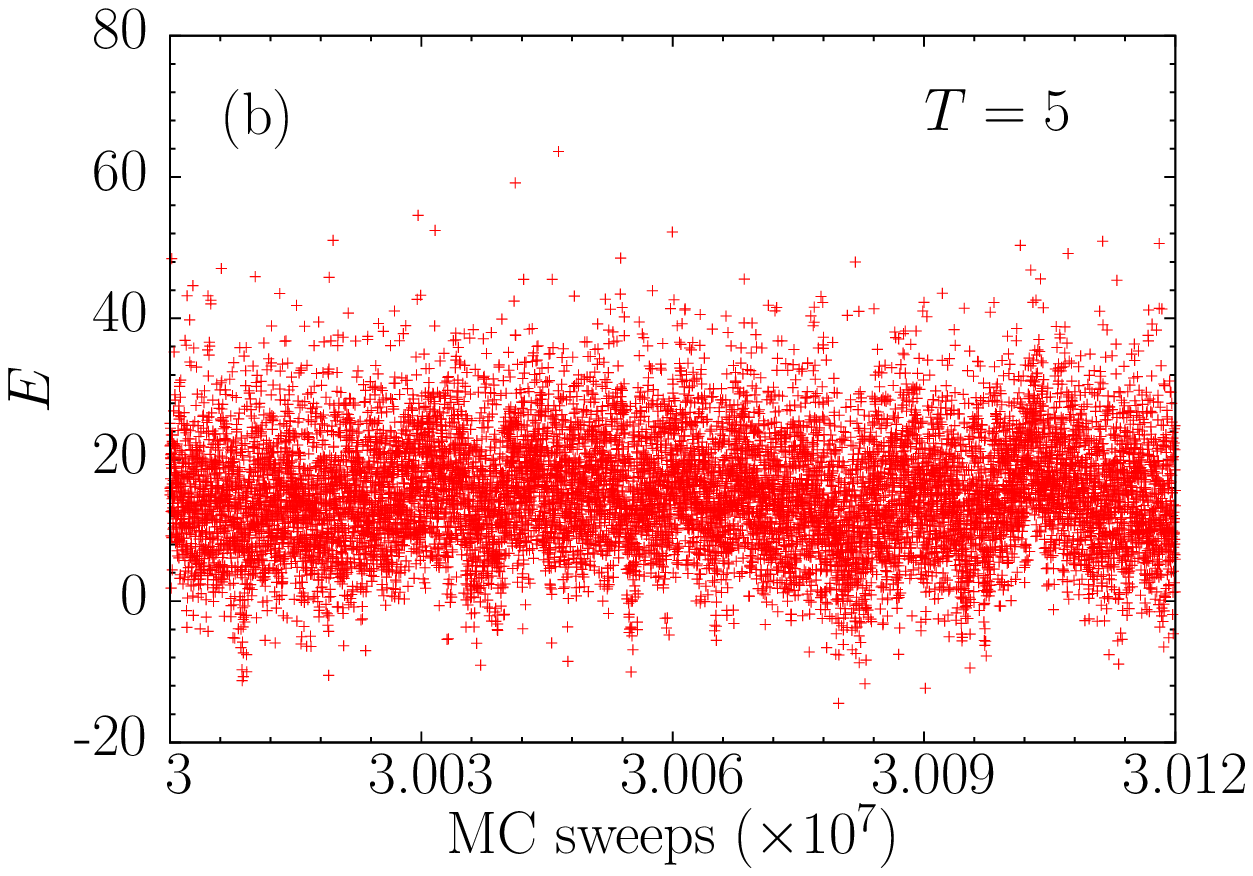}
	\includegraphics[width=5.8cm]{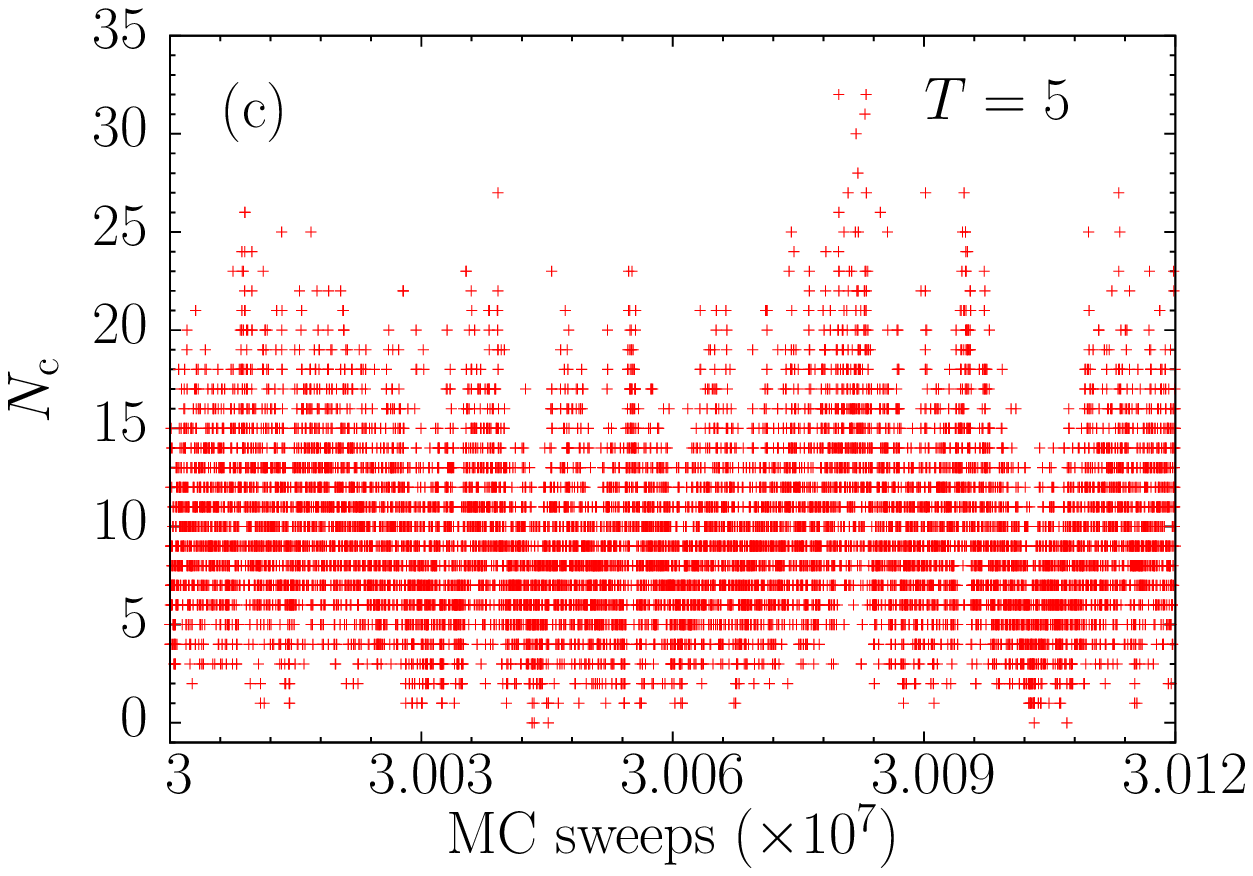} 
	\caption{\label{time_series} (a), (b), and (c) are parts of the time
series of $R^2_{\mathrm{gyr}}$, $E$, and $N_{\mathrm{c}}$ at $T=5$ in
equilibrium for the 30-mer.}
\end{figure*}

It is necessary to mention that when using the binning method to calculate
error bars one needs to ensure that the binning block length is much larger
than the autocorrelation time. The reason is obvious from
Fig.~\ref{auto_30_seperate}. If the autocorrelation time estimated by the
binning method has not yet converged, the estimate
$\tilde{\tau}_{O,\text{bin}}$ is less than the integrated autocorrelation time
($\tilde{\tau}_{O,\text{bin}} < \tau_{O,\text{int}}$). Therefore, the
estimated standard deviation 
\begin{equation}
	\epsilon^2_{\overline{O}} = \frac{\tilde{\sigma}^2_{\overline{O}^{\mathrm{B}},\mathrm{c}}}{K} = \frac{2\tilde{\sigma}^2_O}{N}\tilde{\tau}_{O,\text{bin}} \label{error_bin_2} 
\end{equation}
underestimates the true
value $\epsilon^2_{\overline{O}}=2\sigma^2_{O}\tau_{O,\text{int}}/N$ in this
case, yielding a too small error estimate. 

After the preliminary
considerations, we will now discuss how the dependence of the autocorrelation
time on the temperature can be utilized for the identification of structural
transitions in the polymer system.

\subsection{Slowing down at the $\Theta$ point}

\begin{figure*}
	\centering
	\includegraphics[width=7.8cm]{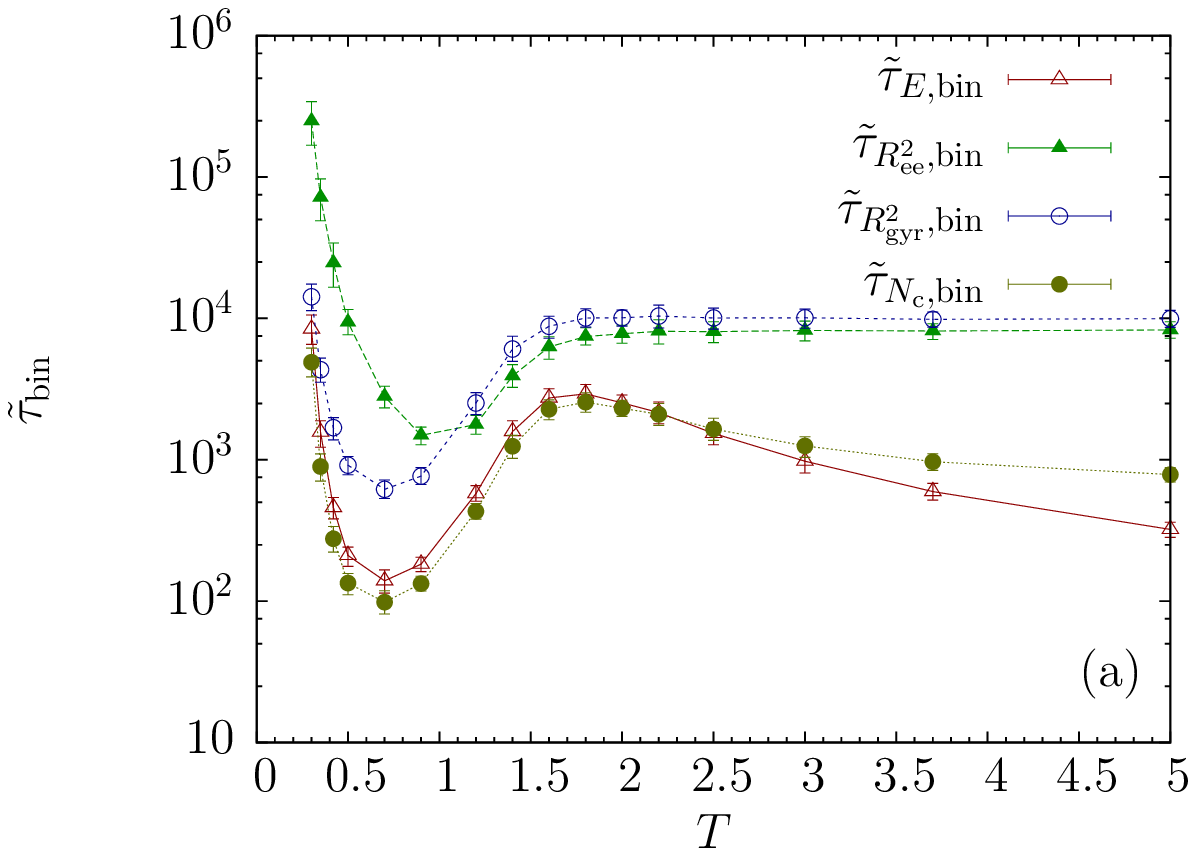}
	\includegraphics[width=7.8cm]{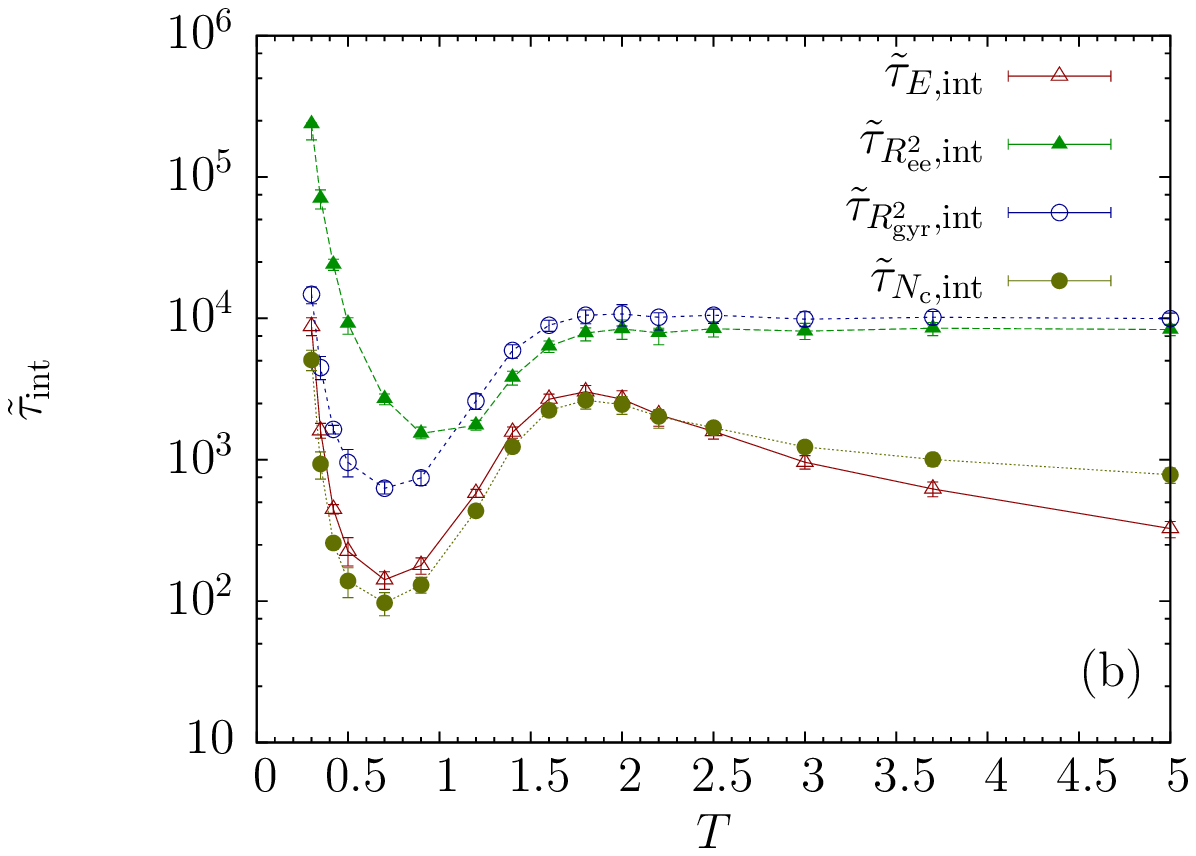} 
	\caption{\label{auto_55} Same as Fig.~\ref{auto_30}, but for the
55-mer.}
\end{figure*}

Figure~\ref{auto_30} shows how the fitted estimated integrated
autocorrelation times $\tau^f_{O}$ vary with temperature. As the comparison
shows, the autocorrelation times estimated by using the binning analysis are
in very good agreement with the results obtained by integrating
autocorrelation functions.

The integrated autocorrelation time curves of $R^2_{\mathrm{ee}}$
and $R^2_{\mathrm{gyr}}$ behave similarly at most of the temperatures except
the temperatures close to the freezing transition. This is not surprising as
both are structural quantities that are defined to describe the compactness of
the polymer. In addition, the integrated autocorrelation time curves of $E$
and $N_{\mathrm{c}}$ behave similarly. Their relation can be understood as
following. The polymer conformation in the solid phase is characterized by
locally crystalline or amorphous metastable structures. Therefore, the main
contribution of each monomer to the energy in this phase originates from the
interaction between this monomer and its non-bonded nearest neighbors. This is
also reflected by the number of contacts  to the nearest neighbors. Thus,
$E\propto N_{\mathrm{c}}$ in the solid phase (see Fig.~1(a)). The
autocorrelation times of the two structural quantities are always larger than
the ones of $E$ and $N_{\mathrm{c}}$. The reason 
is that these quantities are not particularly sensitive to
conformational changes within a single phase. Furthermore, the displacement
update used here does not allow for immediate substantial changes. This can be
seen in Fig.~\ref{time_series}(a) where the time series are shown at high
temperature. From Fig.~\ref{time_series}(b) and \ref{time_series}(c), one
notices that $E$ and $N_{\mathrm{c}}$ fluctuate more strongly than
$R^2_{\mathrm{gyr}}$. 

The most important observation from Fig.~\ref{auto_30} is that slowing down
appears near $T\approx1.4$ which signals the collapse transition. This
temperature is close to the peak positions of the structural fluctuations
shown in Fig.~\ref{canonical_quantities}(d). Within this temperature region,
the autocorrelation time becomes extremal. Large parts of the polymer have to
behave cooperatively which slows down the overall collapse dynamics.

Near the freezing transition ($T\approx0.3$), the autocorrelation times of all
four quantities rapidly increase. Since Metropolis simulations with local
updates typically get stuck in metastable states of the polymer at low
temperatures, we do not estimate autocorrelation times in the $T<0.26$ region.
The freezing transition is, therefore, virtually inaccessible to any
autocorrelation time analysis based on local-update Metropolis simulations.
This is amplified by the fact that the autocorrelation time increase naturally
at low temperatures, because of the low entropy. That means if there would be
a signal of the freezing transition at all in the autocorrelation time curves,
it would be difficult to identify it.

The autocorrelation times of $R^2_{\mathrm{ee}}$, $R^2_{\mathrm{gyr}}$, and
$N_{\mathrm{c}}$ seem to converge to constant values at high temperatures,
whereas the autocorrelation time of $E$ decays. This is partly due to the fact
that the structural quantities and $N_{\mathrm{c}}$ possess upper limiting
values that are reached at high temperatures, thereby reducing the fluctuation
width at constant displacement range. This is a particular feature of the
results obtained in simulations with fixed maximum displacement and it is
different if the acceptance rate
is kept constant instead. This will be discussed in Sec.~\ref{auto fixed accp rate}.

The overall behavior is similar to Metropolis dynamics for the
two-dimensional Ising model on the square lattice, in which the external field
is excluded so that $E\in[-2JL^2,2JL^2]$ where $J>0$ is the coupling constant
and $L$ is the lattice size and $M\in[-L^2,L^2]$ \cite{NewmanMC1999}.

In order to verify that the general autocorrelation properties apply also to
larger polymers, we repeated the simulations for a 55-mer. From
Fig.~\ref{auto_55}, we notice that the behavior is qualitatively the same, but
the autocorrelation times of all quantities are larger than the ones for the
30-mer, as expected. This supports our hypothesis that the qualitative
behavior of the autocorrelation times of the 30-mer is generic and
representative for autocorrelation properties of larger polymers. In
particular, this method offers a possible way for the identification of
transitions, where standard canonical analysis of quantities such as the
specific heat fails.


\subsection{Autocorrelation times at a fixed acceptance rate}
\label{auto fixed accp rate}
\begin{figure*}
	\centering
	\includegraphics[width=15.3cm]{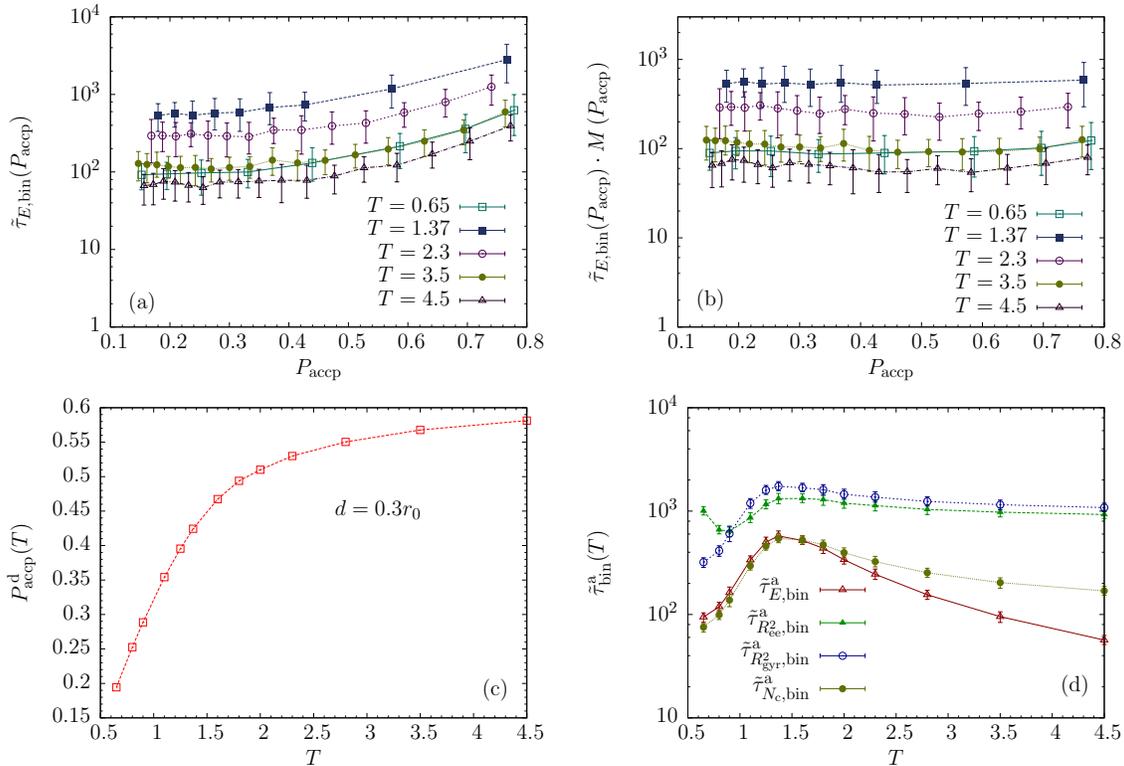}
	\caption{\label{fixed_accp}(a) Integrated autocorrelation times of
the energy $\tilde{\tau}_{E,\text{bin}}$ at different acceptance rates
$P_{\text{accp}}$ for various temperatures near the collapse transition of the
30-mer; (b) modified autocorrelation times of energy versus acceptance rates;
(c) acceptance rates at fixed maximum displacement $d=0.3r_0$ 
for different temperatures; (d) integrated autocorrelation times at fixed 
acceptance rate $P^{\mathrm{\,a}}_{\mathrm{accp}}=0.2$ as a function of temperature for the 30-mer.}
\end{figure*}

In order to find out how the autocorrelation time changes at a fixed
acceptance rate rather than at a fixed maximum displacement range, we used the
binning method to calculate the integrated autocorrelation times at different
constant acceptance rates $P_{\text{accp}}$ for different temperatures near
the collapse transition for the 30-mer. The results for the energetic
autocorrelation times $\tilde{\tau}_{E,\mathrm{bin}}(P_{\mathrm{accp}})$
are shown in Fig.~\ref{fixed_accp}(a), measured
for five different temperatures. Autocorrelation times of the other
quantities exhibit a similar behavior. Two important conclusions can be drawn:
(i) the values of the autocorrelation times depend on acceptance rate and temperature,
but (ii) the monotonic behavior of $\tilde{\tau}_{E,\mathrm{bin}}$ as
a function of $P_{\mathrm{accp}}$ is virtually independent of the temperature. Thus,
if multiplied by a temperature-independent
empirical modification factor
\begin{equation}
	M(P_{\text{accp}}) = e^{-4|P_{\text{accp}}-0.2|^{1.65}}, \label{Multi_factor}
\end{equation} 
the modified autocorrelation time curves become almost independent
of $P_{\mathrm{accp}}$ at these temperatures (see Fig.~\ref{fixed_accp}(b)):
\begin{equation}
	\tilde{\tau}_{\text{bin}}(P_{\mathrm{accp}}) \cdot M(P_{\mathrm{accp}}) \approx \mathrm{const.} \label{tau_fix_accp}
\end{equation} 
in the interval $P_{\mathrm{accp}}\in(0.14,0.78)$. This feature of uniformity
in monotonic behavior and the empirical modification factor \eqref{Multi_factor}
can then be used to modify the autocorrelation times at all temperatures.
For this purpose one reads the autocorrelation
time $\tilde{\tau}^{\mathrm{d}}_{\mathrm{bin}}(T)$
and the acceptance rate $P^{\mathrm{\,d}}_{\mathrm{accp}}(T)$
at fixed maximum displacement at a given temperature $T$ from 
Fig.~\ref{auto_30}(a) and Fig.~\ref{fixed_accp}(c), respectively,
calculates the modification factor $M(P^{\mathrm{\,d}}_{\mathrm{accp}}(T))$
from Eq.~\eqref{Multi_factor}, and obtains the modified autocorrelation
time $\tilde{\tau}^{\mathrm{a}}_{\mathrm{bin}}$ at constant acceptance rate
$P^{\mathrm{\,a}}_{\mathrm{accp}}$ by making use of Eq.~\eqref{tau_fix_accp}.
For simplicity, we choose $P^{\mathrm{\,a}}_{\mathrm{accp}}=0.2$, which yields
\begin{equation}
	\tilde{\tau}^{\mathrm{a}}_{\text{bin}}(T) = \tilde{\tau}^{\mathrm{d}}_{\text{bin}} (T) \cdot M(P^{\mathrm{\,d}}_{\mathrm{accp}}(T)), \label{tau_fix_accp_2}
\end{equation} 
since $M(0.2)=1$. The temperature dependence of this modified
autocorrelation time is shown
in Fig.~\ref{fixed_accp}(d). One notices that the peaks indicating the
 collapse transition are more pronounced than the ones in the fixed maximum
 displacement case, but qualitatively (and quantitatively regarding the
$\Theta$ transition point) this modified approach leads to similar results.
In the temperature range investigated here, the autocorrelation times of all
quantities seem to decrease above the $\Theta$ point. This is different than
the behavior at fixed
maximum displacement range (cp. Fig.~\ref{auto_30}(a)).


\section{SUMMARY}
\label{summary}

Employing the Metropolis Monte Carlo algorithm, we have performed computer
simulations of a simple coarse-grained model for flexible, elastic polymers to
investigate the autocorrelation time properties for different quantities. Two
different methods were employed to estimate autocorrelation times as functions
of temperatures for polymers with 30 and 55 monomers: by integration of
autocorrelation functions and by using the binning method. The results
obtained for different energetic and structural quantities by averaging over
more than 20 independent simulations are consistent.

The major result of our study is that autocorrelation time changes can be used
to locate structural transitions of polymers, because of algorithmic slowing
down. We deliberately employed Metropolis sampling and local displacement
updates, as slowing down is particularly apparent in this case. We could
clearly identify the collapse transition point for the two chain lengths
investigated. Low-temperature transitions are not accessible because of the
limitations of Metropolis sampling in low-entropy regions of the state space.

The identification of transitions by means of autocorrelation time analysis
is, therefore, an alternative and simple method to more advanced technique
such as microcanonical analysis
\cite{MBachmann,Gross2001,Janke1998,Behringer2006,Schnabel2011} or by
investigating partition function zeros
\cite{YangLee,Fisher1965,Janke2001-2002,JRocha}. Those methods require
the precise estimation of the density of states of the system which can only
be achieved in sophisticated generalized-ensemble simulations. The
autocorrelation time analysis method is very robust and can be used as an
alternative method for the quantitative estimation of transition temperatures,
in particular, if the more
qualitative standard canonical analysis of ``peaks'' and ``shoulders'' in
fluctuating quantities remains inconclusive.


\begin{acknowledgments}

The authors thank W. Paul and J. Gross for helpful discussions. This
work has been supported partially by the NSF under Grant No.\ DMR-1207437, and
by CNPq (National Council for Scientific and Technological Development,
Brazil) under Grant No.\ 402091/2012-4.
\end{acknowledgments}



\begin{thebibliography}{99}


\bibitem{Verdier1962}
	P.H. Verdier and W.H. Stockmayer, J. Chem. Phys. \textbf{36}, 227 (1962).

\bibitem{Verdier1966}
	P.H. Verdier, J. Chem. Phys. \textbf{45}, 2122 (1966).

\bibitem{Verdier1972}
	D.E. Kranbuehl and P.H. Verdier, J. Chem. Phys. \textbf{56}, 3145 (1972).


\bibitem{Verdier1973}
	P.H. Verdier, J. Chem. Phys. \textbf{59}, 6119 (1973).

\bibitem{Kranbuehl1973}
	D.E. Kranbuehl, P.H. Verdier, and J.M. Spencer, J. Chem. Phys. \textbf{59}, 3861 (1973).

\bibitem{Kranbuehl1977}
	D.E. Kranbuehl and P.H. Verdier, J. Chem. Phys. \textbf{67}, 361 (1977).


\bibitem{McCormick2005}
	J.A. McCormick, C.K. Hall, and S.A. Khan,  J. Chem. Phys. \textbf{122}, 114902 (2005).



\bibitem{Pestryaev2011}
	E.M. Pestryaev, J. Phys.: Conf. Ser. \textbf{324}, 012031 (2011).


\bibitem{Aoyagi2001}
	T. Aoyagi, J. Takimoto, and M. Doi, J. Chem. Phys. \textbf{115}, 552 (2001).

\bibitem{Malevanets2000}
	A. Malevanets and J.M. Yeomans, Europhys. Lett. \textbf{52}, 231 (2000).

\bibitem{Polson2006}
	J.M. Polson and J.P. Gallant, J. Chem. Phys. \textbf{124}, 184905 (2006).

\bibitem{Bishop1979}
	M. Bishop, M.H. Kalos, and H.L. Frisch, J. Chem. Phys. \textbf{70}, 1299 (1979).

\bibitem{Rapaport1979}
	D.C. Rapaport, J. Chem. Phys. \textbf{71}, 3299 (1979).


\bibitem{Bruns1981}
	W. Bruns and R. Bansal, J. Chem. Phys. \textbf{74}, 2064 (1981).

\bibitem{Bansal1981}
	W. Bruns and R. Bansal, J. Chem. Phys. \textbf{75}, 5149 (1981).

\bibitem{Mussawisade2005}
	K. Mussawisade, M. Ripoll, R.G. Winkler, and G. Gompper, J. Chem. Phys. \textbf{123}, 144905 (2005).


\bibitem{Nidras1997}
	P.P. Nidras and R. Brak, J. Phys. A: Math. Gen. \textbf{30}, 1457 (1997).


\bibitem{Landau2000}
	D.P. Landau and K. Binder, \emph{A Guide to Monte Carlo Simulations in Statistical Physics} (Cambridge University Press, Cambridge, 2000).

\bibitem{NewmanMC1999}
	M.E.J. Newman and G.T. Barkema, \emph{Monte Carlo Methods in Statistical Physics}, (Oxford University Press, Oxford, 1999).

\bibitem{WJanke2002}
	W. Janke, \emph{Statistical Analysis of Simulations: Data Correlations and Error Estimation}, in \emph{Proceedings of the Winter School ``Quantum Simulations of Complex Many-Body Systems: From Theory to Algorithms''}, John von Neumann Institute for Computing, J\"ulich, NIC Series vol. \textbf{10}, ed. by J. Grotendorst, D. Marx and A. Muramatsu (NIC, J\"ulich, 2002), p. 423.

\bibitem{WJanke1996}
	W. Janke, \emph{Monte Carlo Simulations of Spin Systems}, in: \emph{Computational Physics: Selected Methods $-$ Simple Exercises $-$ Serious Applications}, eds. K.H. Hoffmann and M. Schreiber (Springer, Berlin, 1996), p. 10.

\bibitem{WJanke1998}
	W. Janke, \emph{Nonlocal Monte Carlo Algorithms for Statistical Physics Applications}, Mathematics and Computers in Simulations \textbf{47}, 329 (1998).

\bibitem{MetroRosTell1953}
	N. Metropolis, A.W. Rosenbluth, M.N. Rosenbluth, A.H. Teller, and E. Teller, J. Chem. Phys. \textbf{21}, 1087 (1953).



\bibitem{Sokal1989}
	A.D. Sokal, \emph{Monte Carlo Methods in Statistical Mechanics: Foundations and New Algorithms}, lecture notes, Cours de Troisi\`eme Cycle de la Physique en Suisse Romande, Lausanne, 1989.

\bibitem{Sokal1992}
	A.D. Sokal, \emph{Bosonic Algorithms}, in: \emph{Quantum Fields on the Computer}, ed. M. Creutz (World Scientific, Singapore, 1992), p. 211.


\bibitem{Nightingale1996}
	M.P. Nightingale and H.W.J. Bl\"ote, Phys. Rev. Lett. \textbf{76}, 4548 (1996).

\bibitem{Matz1994}
	R. Matz, D.L. Hunter, and N. Jan, J. Stat. Phys. \textbf{74}, 903 (1994).

\bibitem{Coddington1992}
	P.D. Coddington and C.F. Baillie, Phys. Rev. Lett. \textbf{68}, 962 (1992).

\bibitem{Kandel1988}
	D. Kandel, E. Domany, D. Ron, A. Brandt, and E. Loh, Phys. Rev. Lett.
\textbf{60}, 1591 (1988).


\bibitem{Kandel1989}
	D. Kandel, E. Domany, and A. Brandt, Phys. Rev. B \textbf{40}, 330 (1989).


\bibitem{Gunton1983}
	J.D. Gunton, M.S. Miguel, and P.S. Sahni, \emph{The Dynamics of First
Order Phase Transitions} in: \emph{Phase Transitions and Critical Phenomena},
Vol. 8, eds. C. Domb and J.L. Lebowitz (Academic Press, New York, 1983), p.
269.

\bibitem{Binder1987}
	K. Binder, Rep. Prog. Phys. \textbf{50}, 783 (1987).

\bibitem{Herrmann1992}
	H.J. Herrmann, W. Janke, and F. Karsch (eds.), \emph{Dynamics of First Order Phase Transitions} (World Scientific, Singapore, 1992).

\bibitem{JankeFirst-Order1994}
	W. Janke, \emph{Recent Developments in Monte Carlo Simulations of First-Order Phase Transitions}, in: \emph{Computer Simulations in Condensed Matter Physics VII}, eds. D.P. Landau, K.K. Mon and H.-B. Sch\"uttler (Springer, Berlin, 1994), p. 29.

\bibitem{JankePhaseTransitions}
	W. Janke, \emph{First-Order Phase Transitions}, in: \emph{Computer Simulations of Surfaces and Interfaces}, NATO Science Series, II. Mathematics, Physics and Chemistry - Vol. \textbf{114}, Proceedings of the NATO Advanced Study Institute, Albena, Bulgaria, 9 - 20 September 2002, edited by B. D\"unweg, D.P. Landau, and A.I. Milchev (Kluwer, Dordrecht, 2003); pp. 111 - 135.

\bibitem{Berg1991}
	B.A. Berg and T. Neuhaus, Phys. Lett. \textbf{B267}, 249 (1991); Phys. Rev. Lett. \textbf{68}, 9 (1992).


\bibitem{Janke1992}
	W. Janke, B.A. Berg, and M. Katoot, Nucl. Phys. \textbf{B382}, 649 (1992).

\bibitem{Berg1993}
	B.A. Berg, U. Hansmann, and T. Neuhaus, Phys. Rev. B \textbf{47}, 497 (1993); Z. Phys. B \textbf{90}, 229 (1993).


\bibitem{Billoire1993}
	A. Billoire, T. Neuhaus, and B.A. Berg, Nucl. Phys. B \textbf{396}, 779 (1993).

\bibitem{Grossmann1992}
	B. Grossmann and M.L. Laursen, Int. J. Mod. Phys. \textbf{C3}, 1147 (1992); Nucl. Phys. B \textbf{408}, 637 (1993).

\bibitem{Janke1993}
	W. Janke and T. Sauer, Phys. Rev. E \textbf{49}, 3475 (1994).

\bibitem{MBachmann}
	M. Bachmann, \emph{Thermodynamics and Statistical Mechanics of
Macromolecular Systems}, (Cambridge University Press, Cambridge, 2014).

\bibitem{BiCuArHa1987}
	R.B. Bird, C.F. Curtiss, R.C. Armstrong, and O. Hassager, \emph{Dynamics of Polymeric Liquids}, 2nd ed. (Wiley, New York, 1987).


\bibitem{MilBhaBin2001}
 	A. Milchev, A. Bhattacharya, and K. Binder, Macromolecules \textbf{34}, 1881 (2001).

\bibitem{MBachmann2006-1}
	M. Bachmann and W. Janke, Phys. Rev. E \textbf{73}, 020901(R) (2006).

\bibitem{MBachmann2006-2}
	M. Bachmann and W. Janke, Phys. Rev. E \textbf{73}, 041802 (2006).

\bibitem{TVogel2007}
	T. Vogel, M. Bachmann, and W. Janke, Phys. Rev. E \textbf{76}, 061803 (2007).

\bibitem{Gross2001}
	D.H.E. Gross, \emph{Microcanonical Thermodynamics} (World Scientific, Singapore, 2001).

\bibitem{Janke1998}
	W. Janke, Nucl. Phys. B, Proc. Suppl. \textbf{63A-C}, 631 (1998).

\bibitem{Behringer2006}
	H. Behringer and M. Pleimling, Phys. Rev. E \textbf{74}, 011108 (2006).

\bibitem{Schnabel2011}
	S. Schnabel, D.T. Seaton, D.P. Landau, and M. Bachmann, Phys. Rev. E \textbf{84}, 011127 (2011).

\bibitem{YangLee}
	C.N. Yang and T.D. Lee, Phys. Rev. \textbf{87}, 404 (1952); T.D. Lee and C.N. Yang, Phys. Rev. \textbf{87}, 410 (1952).

\bibitem{Fisher1965}
	M.E. Fisher, in \emph{Lectures in Theoretical Physics vol. 7C,} ed. by W.E. Brittin (University of Colorado Press, Boulder, 1965), Chap. 1.

\bibitem{Janke2001-2002}
	W. Janke and R. Kenna, J. Stat. Phys. \textbf{102}, 1211 (2001); Comp. Phys. Comm. \textbf{147}, 443 (2002); Nucl. Phys. B (Proc. Suppl.) \textbf{106-107}, 905 (2002).

\bibitem{JRocha}
	J.C.S. Rocha, S. Schnabel, D.P. Landau, and M. Bachmann, Phys. Rev. E, in press (2014).

\end{thebibliography}
\end{document}